\documentclass[a4paper,fleqn]{cas-dc}

\usepackage[authoryear]{natbib}

\usepackage{threeparttable}
\usepackage{lscape}
\usepackage{multirow}
\usepackage{longtable}
\usepackage{makecell}

\def\tsc#1{\csdef{#1}{\textsc{\lowercase{#1}}\xspace}}
\tsc{WGM}
\tsc{QE}
\tsc{EP}
\tsc{PMS}
\tsc{BEC}
\tsc{DE}

\begin{document}
\let\WriteBookmarks\relax
\def\floatpagepagefraction{1}
\def\textpagefraction{.001}
\shorttitle{accident analysis and preventtion}
\shortauthors{Ke Wang et~al.}

\title [mode = title]{The Adaptability and Challenges of Autonomous Vehicles to Pedestrians in Urban China}

 \author[1]{Ke Wang}
\cormark[1]
\fnmark[1]
 \ead{kw@cqu.edu.cn}

  \author[1]{Gang Li}
\fnmark[1]
 \ead{ligangvehicle@cqu.edu.cn}

 \author[2]{Junlan Chen}
\cormark[1]
 \ead{junlanchen@cqnu.edu.cn}

   \author[1]{Yan Long}
 \ead{longyan@cqu.edu.cn}

 \ead{wang.dongqiang@ciat-cq.org}
 \author[3]{Tao Chen}
  \author[3]{Long Chen}
 
  \author[3]{Qin Xia}

\fntext[fn1]{These authors contributed equally to this work.}
\cortext[cor1]{Corresponding author.}
\address[1]{School of Automobile Engineering, the Key Lab of Mechanical Transmission, Chongqing University, Chongqing 400044, China}
\address[2]{School of Economics \& Management, Chongqing Normal University, Chongqing 401331, China}
\address[3]{State Key Laboratory of vehicle NVH and Safety Technology, China Automotive Engineering Research Institute Company, Ltd., Chongqing 401122, China}

\begin{abstract}
China is the world's largest automotive market and is ambitious for autonomous vehicles (AVs) development. As one of the key goals of AVs, pedestrian safety is an important issue in China. Despite the rapid development of driverless technologies in recent years, there is a lack of researches on the adaptability of AVs to pedestrians.  To fill the gap, this study would discuss the adaptability of current driverless technologies to China urban pedestrians by reviewing the latest researches. The paper firstly analyzed typical Chinese pedestrian behaviors and summarized the safety demands of pedestrians for AVs through articles and open database data, which are worked as the evaluation criteria. Then, corresponding driverless technologies are carefully reviewed. Finally, the adaptability would be given combining the above analyses. Our review found that autonomous vehicles have trouble in the occluded pedestrian environment and Chinese pedestrians do not accept AVs well. And more explorations should be conducted on standard human-machine interaction, interaction information overload avoidance, occluded pedestrians detection and nation-based receptivity research. The conclusions are very useful for motor corporations and driverless car researchers to place more attention on the complexity of the Chinese pedestrian environment, for transportation experts to protect pedestrian safety in the context of AVs, and for governors to think about making new pedestrians policies to welcome the upcoming driverless cars.     
\end{abstract}


\maketitle

\section{Introduction}
\label{sec:introduction}
\begin{figure}[pos=!t]
\centering
\includegraphics[width=3.5in]{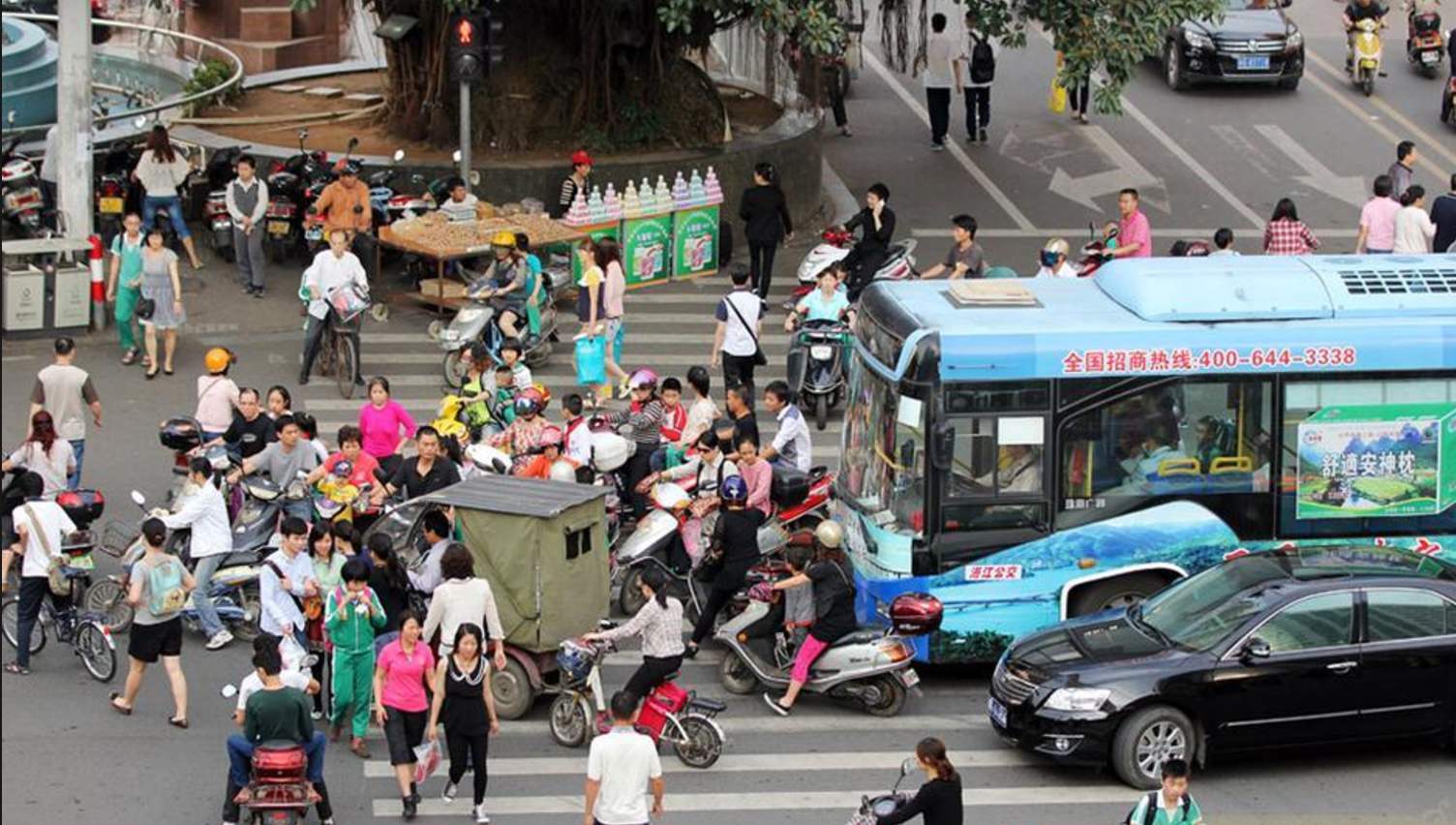}
\DeclareGraphicsExtensions.
\caption{The challgening Chinese pedestrian environment}
\label{figure 1}
\end{figure}

{P}{edestrians} are the vulnerable road users and account for 23\% of world traffic deaths in 2018 according to WHO\citep{RN261}. Naturally, China has the most pedestrians around the world due to its largest population and traffic developing state, and pedestrian safety has always been a problem of China since 26.1\% of traffic deaths are pedestrians in 2013, while in America it is 16.1\%\citep{RN180}. This terrible result comes from bad road behaviors of Chinese pedestrians, such as red light running; a report(N=31649) showed 18.54\% of pedestrians would run red lights in Changsha\citep{RN263}. These bad behaviors make the Chinese traffic environment very challenging. Fig.\ref{figure 1} is the common scene at Chinese crosswalks.  \\

    AVs are deemed as promising solutions for safer road transportation in the future\citep{RN94,RN6998} and China is expected to become one of the largest markets for AVs\citep{RN14}. Therefore, it will be meaningful to analyze the adaptability of AVs to better protect pedestrians and give corporations and governments some useful guidance.\\
    
    Driverless technologies are developing rapidly and according to the ability of self-driving,the Society of Automotive Engineers(SAE) divides self-driving into six levels(L0-L5: from no automation to full automation)\citep{RN288}, corresponding to driving in pedestrians traffic of different difficulty levels. However, the adaptability of driverless technologies to pedestrians is largely unknown. In previous researches, the manuscripts paid attention to pedestrian detection\citep{RN127, RN146,Guofa2019Deep}, interaction\citep{RN478}, receptivity\citep{RN74, RN78}, behavior prediction\citep{RN479}, pose estimation\citep{RN480,RN7668}, etc. These researches are related to the adaptability but did not directly point out the adaptability. \\
    
Similarly, review articles focused more on technology as well. Generally, pedestrian detection is reviewed to summarize and compare the detection algorithms according to the used sensors \citep{RN470, RN469, RN465, RN468, RN472, RN474}. Additionally, Deb et al. summarized the factors that influence the pedestrians' behaviors,  public acceptance of fully automated vehicles as well as current interacting interfaces between pedestrians and autonomous vehicles\citep{RN464}. Cao et al. conducted a review on different methods to model crowd of pedestrians\citep{RN466}. Daniela et al.  explored the ways pedestrians' intention estimation has been studied, evaluated, and evolved\citep{RN467}. Kardi et al. provided a review of a microscopic pedestrian simulation model\citep{RN475}. Sarker et al. reviewed human factors that would influence the acceptance of users to AVs including user comfort, trust, reliability, and preferences \citep{RN497}. In these reviews, the detection methods, influencing factors, and interacting interfaces were carefully summarized and these are closely related to adaptability. Nevertheless, there is a lack of systematic and direct analyses of the adaptability to pedestrians.Therefore, the adaptability analyses are valuable and needs to be supplemented. \\

We have three contributions: firstly, the paper is the first to summarize the Chinese pedestrians' phenomena through abundant data and comparision with foreign countries and analyze the key safety demands for AVs. Secondly, we conducted a comprehensive literature review on the newest driverless technologies to pedestrians, including detection, interaction as well as receptivity. Thirdly, we executed the first try on the adaptability of AVs to Chinese pedestrians and summed up the challenges as well as opportunities. The paper is useful for driverless researchers who care about pedestrian safety and researchers from other areas, such as traffic safety and public policy, who want to conduct researches related to AVs, for we offer them guidance and research directions.   \\

The rest of the paper is organized as follows. Section 2 show the methods that we review and conduct our analyses. Section 3 analyzed three typical phenomena of Chinese pedestrians. A comprehensive literature review of driverless technologies for pedestrians and adaptability analyses were provided in Section 4. In Section 5, emerging challenges and opportunities for future pedestrian research were proposed. Section 6 is the conclusion of this paper.


\section{Methods}
The objective of this article is to figure out the adaptability of AVs to Chinese pedestrians by reviewing articles. The first question should be how to evaluate the adaptability logically. To analyze the adaptability step by step, we divided it into two main questions: what are the Chinese pedestrians' characteristics and their technical demands for AVs, and what are the current abilities of AVs when facing pedestrians? By solving the first, we know the safety demands for Chinese pedestrians, which would be worked as our criteria to evaluate adaptability. Then, through reviewing articles, the current abilities of AVs would be summarized. By combining the two answers, the adaptability would be analyzed and develop the answer to the adaptability of AVs to Chinese urban pedestrians. The methodology is shown in figure 2.\\

Firstly, pedestrain behaviors data was collected from the Chinese government, research articles and websites as the left side of figure 2 showed. In the searching process,  we used terms like 'Chinese urban pedestrians', 'pedestrian bad behaviors', 'pedestrian accidents', 'pedestrian safety' to search results in engine (baidu, baidu scholar, google, google scholar) and filter them by reading the abstract. After collecting data, some typical behaviors of Chinese pedestrians are gained. Then, based on our technical background of AVs, the influence of these pedestrian behaviors on AVs would be carefully discussed. More importantly, some technical demands for AVs are proposed to protect pedestrian safety. In our research, these demands are worked as the criteria to evaluate the adaptability. To make it concrete, we divide it into three classes: excellent, OK, bad, to evaluate the adaptability.   \\

Secondly, according to the demands, corresponding driverless technologies are searched and summarized through scholar engine and authoritative websites. In the searching process, we combined terms like 'autonomous vehicles', 'driverless technologies', 'pedestrians', and 'pedestrian safety' to search professional articles and filtered them according to our demands. After that, the abilities of AVs are categorized into groups and summarized by tables and figures.  Considering the demands, we evaluate the abilities of AVs respectively.  \textbf{One important thing} is that our reviewed technologies are the newest researches and are hard to tell which autonomous level they belong to. Therefore, in this article, we discuss the adaptability based on single technology rather than autonomous levels. Finally, the adaptability summary of every single technology would be our analyses of adaptability. \\

Thirdly, during the process, we found some challenging but promising problems and we analyzed them in a simple way after the adaptability analyses, to give some guidance to practitioners.\\
\begin{figure*}[pos=!t]
\centering
\includegraphics[width=7in]{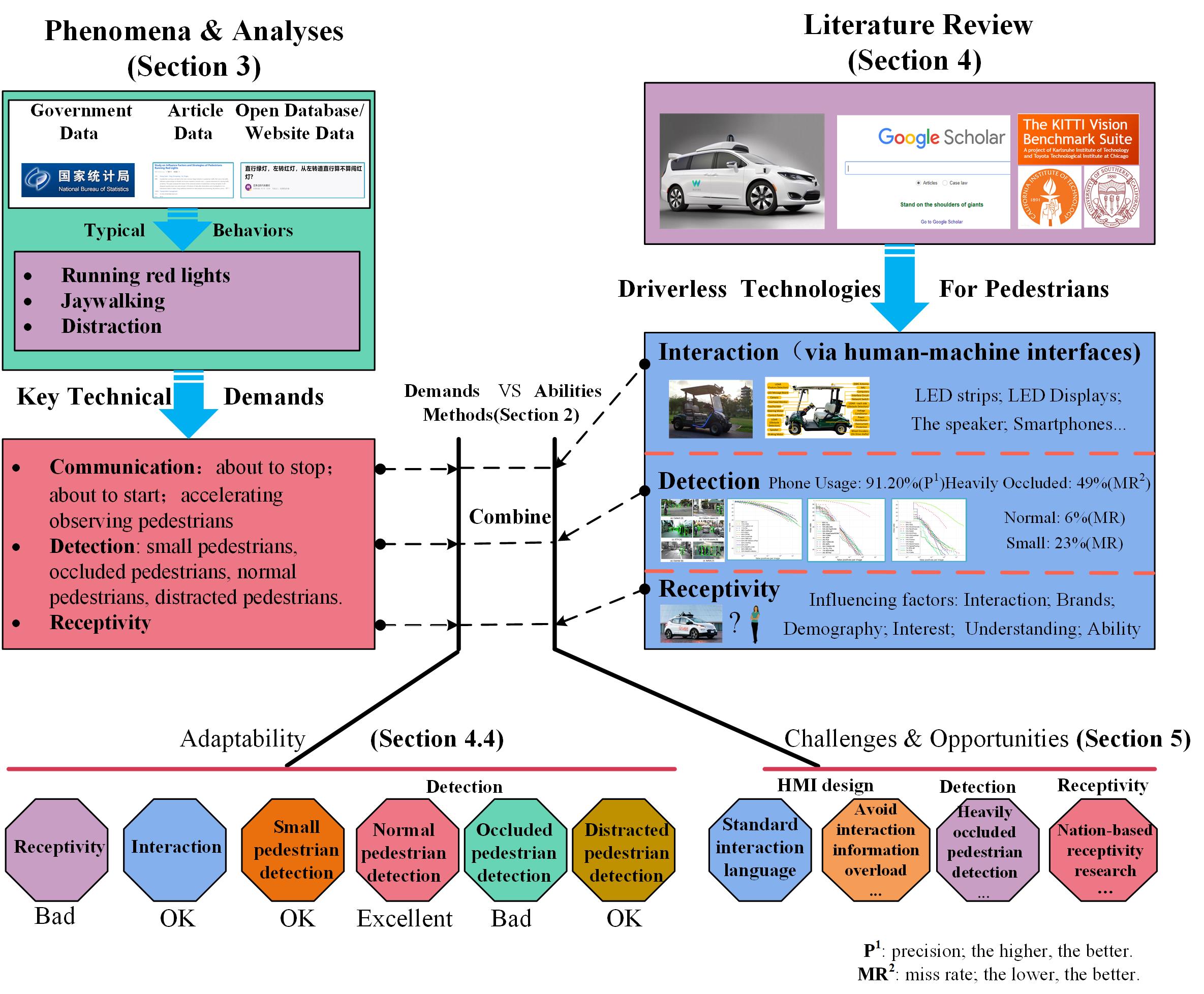}
\DeclareGraphicsExtensions.
\caption{The content structure of this paper}
\label{fig.2}
\end{figure*}

\section{Phenomena \& Analyses}
Pedestrians are vulnerable road users around the world and recognized as the worst victims because they are directly exposed to the impact of traffic crashes compared to vehicle passengers\citep{RN282,RN261}.
In developed countries, pedestrians would be safer. According to Road trauma Australia: 2018 statistical summary\citep{RN269}, the deaths-to-injuries rate of Australian pedestrians is 4\%. In America, 5,977 pedestrians were killed with 71,000 injuries in 2017 according to road traffic statistics\citep{RN267}. While in England, one fatal accident would take place in 50 accidents\citep{RN278}. However, as the biggest developing country, China has more severe pedestrian safety problems. According to the statistics of National Bureau of Statistics of China\citep{RN454}, the deaths-to-accidents ratio of pedestrians is about 50\%, which is far worse than that of car passengers as well as cyclists as Table \ref{table 1} shows; in other words, there will be a person dead in every two pedestrian accidents. Therefore, protecting pedestrians is an essential problem of Chinese traffic safety. \\

\begin{table*}[]
\centering
\caption{The ratio of deaths to accidents of Chinese traffic participants\citep{RN454}}
\label{table 1}
\begin{tabular}{c|ccccccccc} \hline
Traffic participants         & data             & 2017    & 2016    & 2015    & 2014    & 2013    & 2012    & 2011    & 2010    \\\hline
\multirow{3}{*}{All}         & Accidents        & 203049  & 212846  & 187781  & 196812  & 198394  & 204196  & 210812  & 219521  \\
                             & Deaths           & 63772   & 63093   & 58022   & 58523   & 58539   & 59997   & 62387   & 65225   \\
                             & \textbf{Deaths/accidents} & \textbf{0.31} & \textbf{0.30} & \textbf{0.31} & \textbf{0.30} & \textbf{0.30} & \textbf{0.30} & \textbf{0.29} & \textbf{0.30}\\\hline
\multirow{3}{*}{Cyclists}    & Accidents        & 1576    & 1460    & 1369    & 1393    & 1304    & 1433    & 1522    & 1978    \\
                             & Deaths           & 350     & 341     & 304     & 289     & 300     & 279     & 315     & 447     \\
                             & \textbf{Deaths/accidents} & \textbf{0.22} & \textbf{0.23} & \textbf{0.22} & \textbf{0.21} & \textbf{0.23} & \textbf{0.19} & \textbf{0.21} & \textbf{0.23} \\\hline
\multirow{3}{*}{Car passengers}        & Accidents        & 139412  & 145820  & 129155  & 136386  & 138113  & 142995  & 145338  & 148367  \\
                             & Deaths           & 46817   & 45990   & 42388   & 42847   & 42927   & 44679   & 46100   & 46878   \\
                             & \textbf{Deaths/accidents} & \textbf{0.34} & \textbf{0.32} & \textbf{0.33} & \textbf{0.31} & \textbf{0.31} & \textbf{0.31} & \textbf{0.31} & \textbf{0.32} \\\hline
\multirow{3}{*}{Pedestrians} & Accidents         & 2470    & 2443    & 2137    & 2242    & 2088    & 2063    & 2277    & 2565    \\
                             & Deaths           & 1322    & 1304    & 1192    & 1247    & 1185    & 1075    & 1134    & 1222    \\
                             & \textbf{Deaths/accidents} & \textbf{0.54} & \textbf{0.53} & \textbf{0.56} & \textbf{0.56} & \textbf{0.57} & \textbf{0.52} & \textbf{0.50} & \textbf{0.48}  \\\hline
\end{tabular}
\end{table*}

\begin{table}[]
\centering
\caption{Statistics of red lights running of pedestrians in Hangzhou (the first column is the different possibility  to run red lights and the second is corresponding percentages of pedestrians, N=200) \citep{RN457}}
\label{table 2}
\begin{tabular}{ccccc}\hline
Possibility of running red light & 80\%  & 50\%   & 10\% & 0\% \\\hline
Percentage                       & 2.5\% & 17.5\% & 88\% & 1\% \\\hline
\end{tabular}
\end{table}

\begin{table}[]
\centering
\caption{The impact of groups on running red lights of pedestrians in Yulin(Single denotes pedestrians run red lights alone and group means pedestrians together run reds lights, N=3460) \citep{RN458})}
\label{table 3}
\begin{tabular}{m{3cm}<{\centering}m{2cm}<{\centering}m{2cm}<{\centering}}\hline
Pedestrians state & Obey the rules & Run red lights \\\hline
Single                & 76.7\%         & 23.3\%         \\
Group                 & 70.2\%         & 29.8\%        \\\hline
\end{tabular}
\end{table}
\begin{table}

\centering
\caption{Pedestrians’ attitudes toward running red lights in Xian(N=425)\citep{RN459}}
\label{table 4}
\begin{tabular}{m{2cm}<{\centering}m{3cm}<{\centering}m{2cm}<{\centering}}\hline
Attitudes   & Agree to run red lights & Disagree \\\hline
Percentage & 60\%                    & 40\%    \\\hline
\end{tabular}
\end{table}
However, due to the low average education level and awareness of obeying rules, Chinese pedestrians often violate traffic rules and cause accidents, adding complexity to the Chinese driving environment. 
The driverless car is a potential solution to promote traffic safety, however, whether AVs are suitable for the Chinese pedestrian environment is largely unknown. To assess the adaptability, three typical behaviors of Chinese pedestrians were summarized in the following through literature as well as open databases and we found China is one of serious countries of these behaviors. After that, the characteristics of Chinese pedestrians were analyzed. On the basis, the key technical demands for AVs were put forward as the evaluating criteria of adaptability. The followings are the analyses of three typical behaviors.   \\

\begin{figure*}[pos=!t]
\centering
\includegraphics[width=7in]{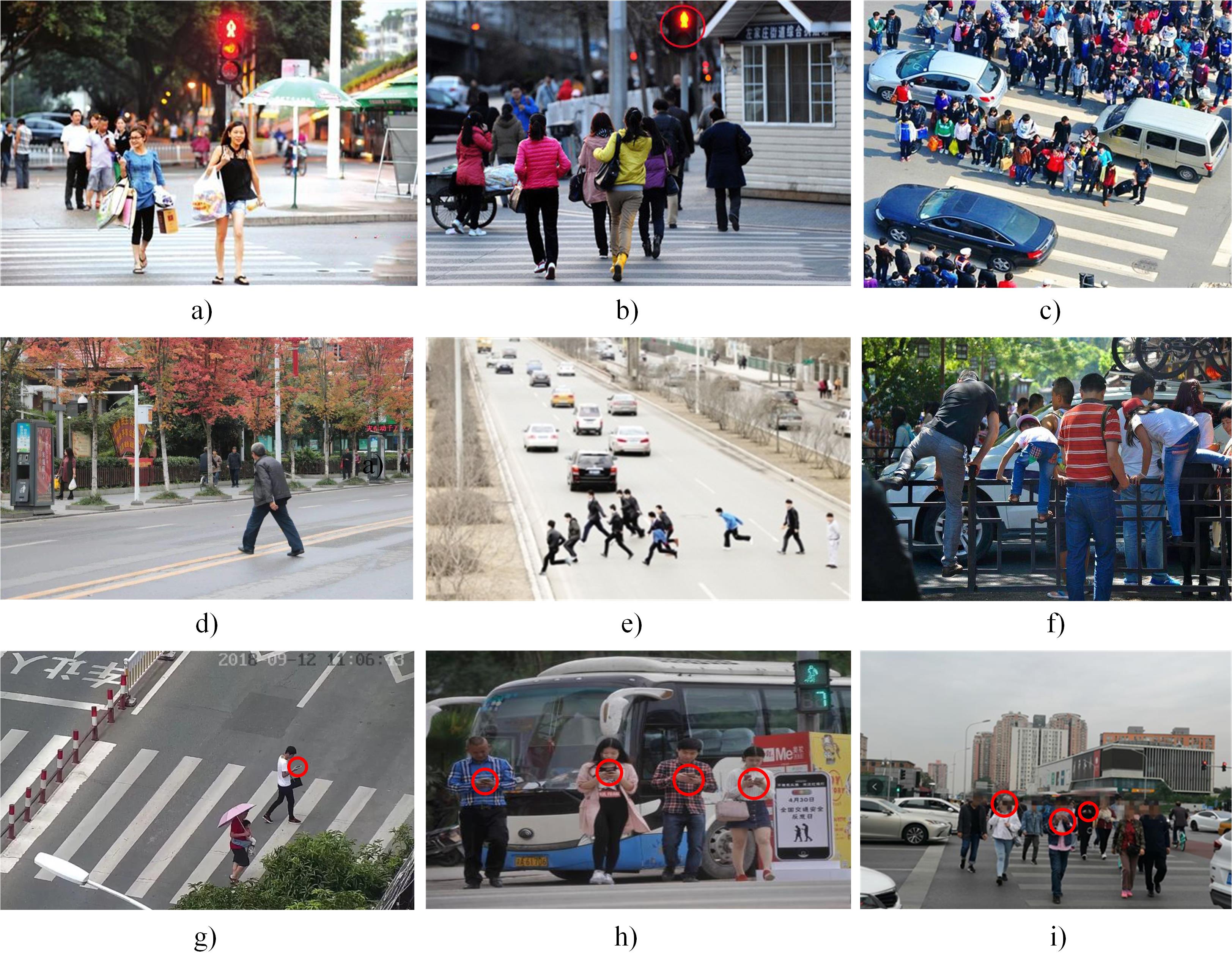}
\DeclareGraphicsExtensions.
\caption{Three typical bad behaviors of Chinese pedestrians: a), b), c) running red lights; d), e), f) jaywalking; g), h), i) distraction (phone utilization)}
\label{fig.3}
\end{figure*}

\subsection{Red lights running}
Red lights running of pedestrians greatly influences their safety. In Lille of France, one third of pedestrians are reported running red lights in 2015\citep{RN265}. In America, 4.5\% of pedestrians traffic deaths happened when they failed to obey traffic signals according to national statistics in 2017\citep{RN267}. However, running lights running is more serious in China and is nicknamed the Chinese style of road crossing. Table \ref{table 2}, \ref{table 3}, \ref{table 4} as well as \ref{table 5} are statistics of running red lights done in four main cities in China. From these data, running red lights is a regular behavior when pedestrians are crossing the crosswalk with a possibility up to 70\%. What is more, the attitudes of pedestrians toward obeying light rules are negative because 60\% of pedestrians are approved of running red lights as table \ref{table 4} shows. Table \ref{table 3} reflects that the number of pedestrians influences the violating behaviors since walking in group increases the possibility to violate lights, which is related to the conformity phenomenon of psychology.\\

\begin{table*}[]
\centering
\caption{Frequency of running red lights of pedestrians in Shanghai(N=500)\citep{RN456}}
\label{table 5}
\begin{tabular}{m{2cm}<{\centering}m{2cm}<{\centering}m{2cm}<{\centering}m{2cm}<{\centering}m{2cm}<{\centering}m{2cm}<{\centering}m{2cm}<{\centering}}\hline
Behaviors  & \multicolumn{3}{c}{Run red lights} & \multicolumn{3}{c}{See others run red lights} \\\hline
Frequency  & Usually  & Occalisonally  & Never  & Usually      & Occasionally      & Never      \\
Percentage & 12\%     & 65\%           & 23\%   & 67\%         & 30\%              & 3\%       \\\hline
\end{tabular}
\end{table*}

The Fig. \ref{fig.3}a, b, c are common scenes in Chinese crosswalks, which shows three levels of running red lights according to the pedestrian number. The conflict in the scenes is that cars have to cross the crosswalk but to keep pedestrians safe.  When there are minority persons violating the lights, their bodies, trajectories, tendencies as well as intent are clear and cars could slow down to avoid at sacrifice of efficiency. To make matters worse, influenced by the conformity phenomenon, persons would violate lights in group as \ref{fig.3}b shows, hence leading to pedestrian occlusion. The occlusion would cause a misunderstanding of pedestrian intent and wrong detection of pedestrians, possibly causing accidents. In rush hours, the pedestrian flow is even walking on the vehicle road. In this case, it is difficult for the cars to move forward but wait.\\
Analyzing the behaviors, we conclude detection plays an essential role in red lights running scenarios. Good detection means cars know the right position as well as features of pedestrians, and could predict the intent. Based on detection, cars could decide whether they should yield or keep their movements when pedestrians are running red lights. Additionally, special care would be given to those special pedestrians like the elderly, the disabled, the distracted as well as the wheelman according to the detected features. In the context of traditional cars, drivers could detect almost all pedestrians in the first level of red lights running according to human intelligence. When more pedestrians violate the lights, drivers impossibly detect all pedestrians as a result of limited attention and occlusion.  AVs could understand the world through multiple sensors, so could they detect pedestrians well in the red lights running scenarios? \\

\begin{figure*}[pos=!t]
\centering
\includegraphics[width=7in]{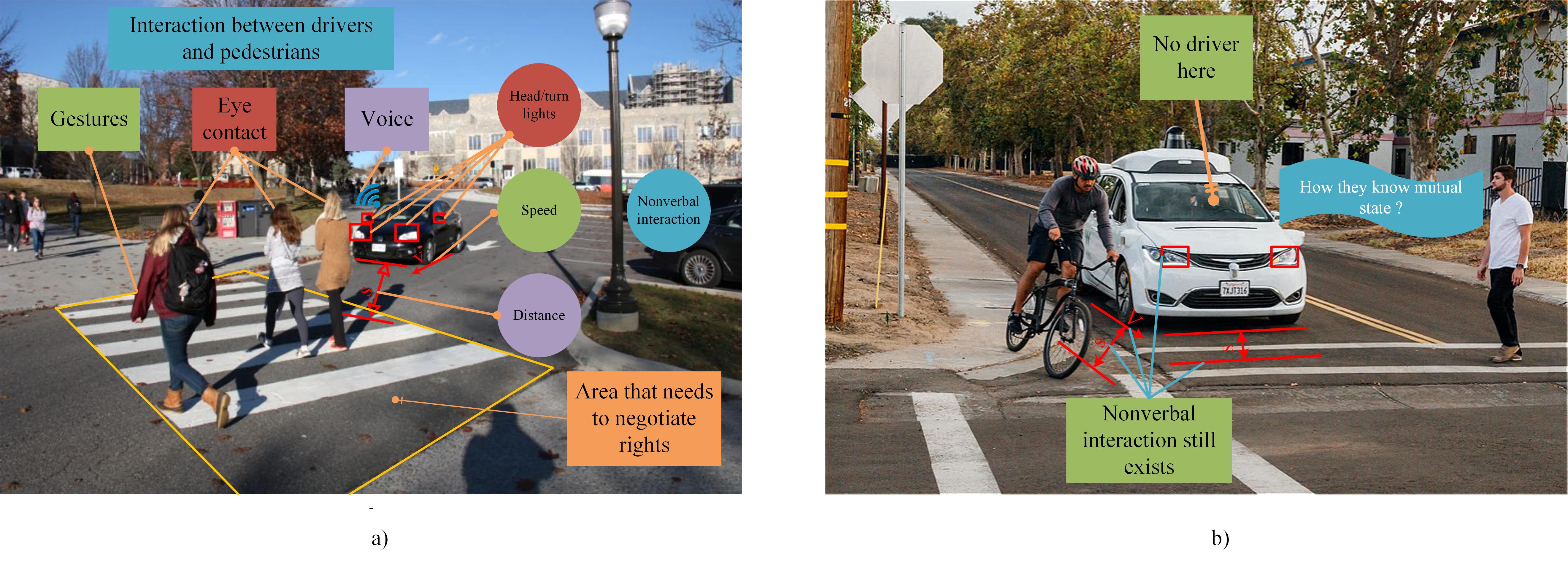}
\DeclareGraphicsExtensions.
\caption{Intent communication methods between pedestrians and vehicles: a) traditional vehicles and pedestrians b) AVs and pedestrians }
\label{fig.4}
\end{figure*}

Additionally, intent communication between pedestrians and vehicles is of vital importance in the process. Generally, intent communication includes the moving state of pedestrians and cars and their observation of each other. Through detection, drivers know pedestrian position and features, however, who should take the road right to go need negotiating, which is tackled by communication.  In the current driving environment, there already exist some nonverbal methods to communicate pedestrians with cars, such as car light (headlights, turn lights, rear lights), distance, voice of the cars. In addition, drivers play an important role in communication. During the process, drivers could interact with pedestrians by utilizing eye contact, gestures as well as voice to assign road right \citep{RN262}as shown in Fig. \ref{fig.4}a,and Lee et al.\citep{RN281} concluded that the existence of drivers would strength the safety-in-numbers effect to protect pedestrians by increasing the interaction between pedestrians and vehicles. More importantly, it will give pedestrians more trust to communicate with humans rather than machines. Compared with traditional cars, there will be no real driver in AVs as Fig. \ref{fig.4}b shows, so how do AVs communicate with pedestrians and whether pedestrians would trust AVs?

\subsection{Jaywalking}
\begin{figure}[pos=!t]
\centering
\includegraphics[width=3.5in]{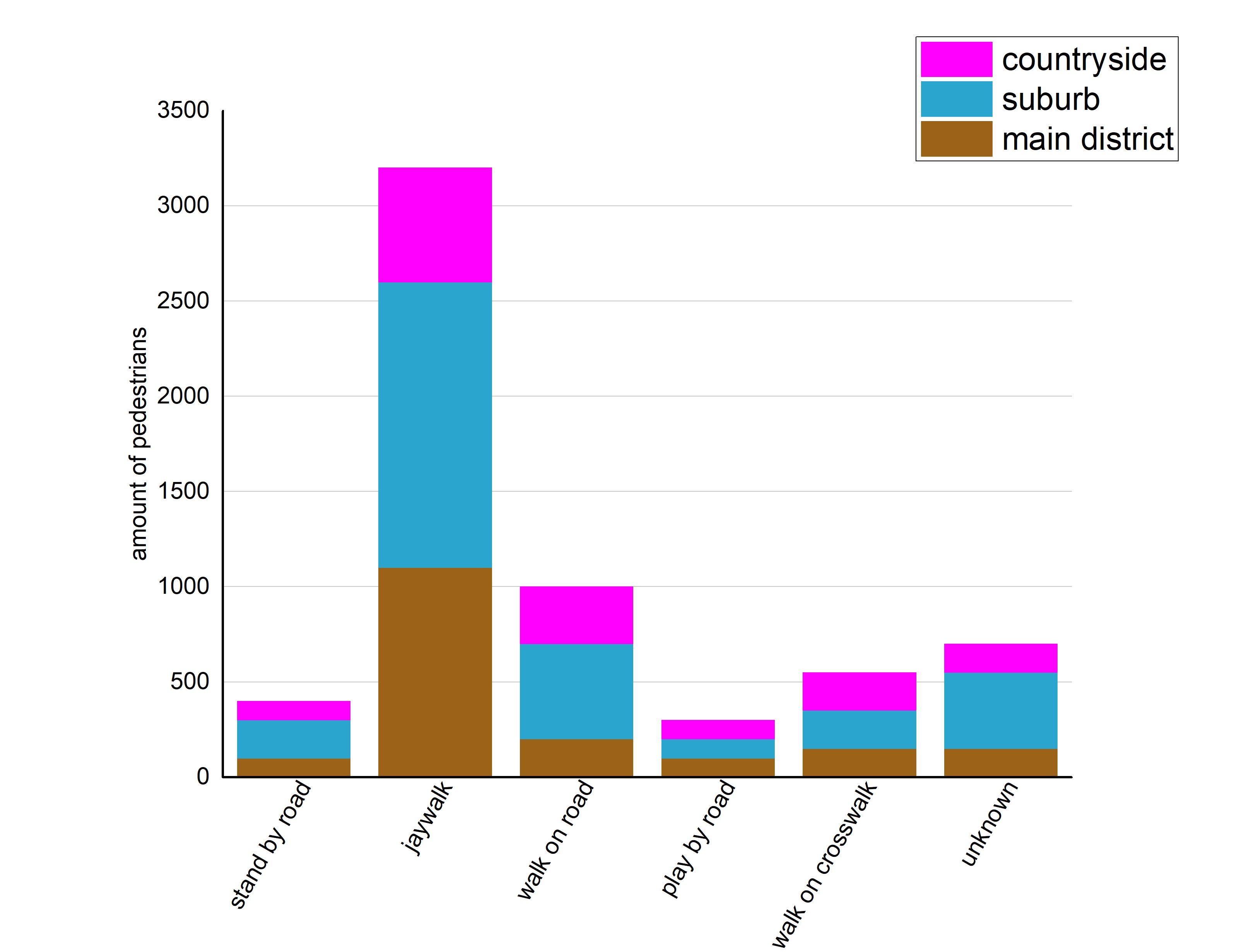}
\DeclareGraphicsExtensions.
\caption{The causes of pedestrian accidents in three areas in Chongqing\citep{RN460}}
\label{fig.5}
\end{figure}

Jaywalking means pedestrians pass a road illegally for convenience where there is no crosswalk or some leading signs, possible with a baluster-passing behaviors. In the UK, nearly 50\% of pedestrians crossed the road ignoring the traffic signals according to YouGov poll 2013\citep{RN270}. In America, a similar YouGov poll in 2014 found that only 13\% of Americans said they had never jaywalked\citep{RN271} and America national statistics showed that 21.2\% of pedestrians traffic deaths came from pedestrians' improper crossing of roadway\citep{RN267}. The Singapore government fined 2,049 jaywalkers in the first quarter of 2013\citep{RN266}. Similarly, jaywalking is very common in China because of unreasonable crosswalk settings and a lack of awareness to obey rules. From Table \ref{table 6}, we could see jaywalking is very dangerous for 68\% pedestrian accidents are related to accidents. What is more, the research of \citep{RN460}showed that no matter in urban, suburbs or countryside, jaywalking is the biggest cause of car accidents according to Fig.\ref{fig.5}. \\
Fig. \ref{fig.3}d, e, f are common scenes of jaywalking in urban China, from which we could observe jaywalking often occurs in the area without signs, even with a fence in the middle of the road. In these scenes, the trajectories of pedestrians are irregular and difficult to predict because there are no signs to regulate their movements, and they are faster than on the crosswalks. For drivers, they would be partly lacking in concentration as a result of no expectation for pedestrians; additionally, cars would be faster due to the higher speed limit on these sections. For the road itself, some special sections, such as curved sections, cars and pedestrians are naturally difficult to detect each other. \\
These features contribute to most of jaywalking accidents, which could be divided into three types. The first type of accidents is the failure to detect each other. The drivers do not observe pedestrians cross the road and hence they crash into pedestrians directly because of no expectation for pedestrians or on special sections. The second is untimely detection to each other. In this case, it is difficult for drivers to avoid a collision as a result of vast inertia out of high speed. The third is poor communication between drivers and pedestrians. With high speed, both drivers and pedestrians have very little time to negotiate the road rights, often leading to poor communication.\\
Similarly, detection is essential to keep pedestrians safe in the jaywalking. In this scenario, cars must detect pedestrians from long distance because cars are so fast that drivers need more time and longer distance to stop. Moreover, enough communication is required to transfer the moving state. With high speed and far distance, the communicating methods between pedestrians and drivers would be limited and easily misunderstood. Therefore, more communicating methods are needed to better transmit the state. Traditional cars are troubled by remote communication and small pedestrian detection. As a high technology, how do AVs interact with remote pedestrians and could they detect remote small pedestrians?

\begin{table*}[]
\centering
\caption{The behaviors of pedestrians just before pedestrian accidents took place(N=181)\citep{RN461}}
\label{table 6}
\begin{tabular}{m{1.5cm}<{\centering}m{1cm}<{\centering}m{2cm}<{\centering}m{2cm}<{\centering}m{1.5cm}<{\centering}m{1.5cm}<{\centering}m{1.5cm}<{\centering}m{1.5cm}<{\centering}}\hline
Behaviors & \textbf{Jaywalk} & Walk on crosswalk in the section & Walk on crosswalk in the intersection & Stand on the road & Work by the road & Walk by the road & others \\\hline
Percentage & \textbf{68\%}       & 5.5\%                               & 7.2\%                                    & 2.8\%                     & 6.6\%            & 7.7\%            & 2.2\% \\\hline
\end{tabular}
\end{table*}

\subsection{Distraction}
Distraction refers to pedestrians doing other things and not paying full attention to the environment when they are passing the road, and the possible reasons are reading, eating, using smartphones, talking, and the influence of alcohol, drugs, or medication. As Judith et al. showed\citep{RN275}, phone usage has been the major reason of pedestrian distraction recently, therefore, the following mainly talk about phone usage.
 \\

\begin{table*}[]
\centering
\caption{The attitudes of using phone on crosswalks of pedestrians in Hefei(questionnaire, N=405)\citep{RN388}}
\label{table 7}
\begin{tabular}{cm{5cm}<{\centering}m{5cm}<{\centering}c}\hline
Attitudes & Use phone when crossing the road this week & Involved in accidents because of phone utilization in crosswalk & Whether to be punished \\\hline
Positive&162(40\%) & 22(5.4\%)& 206(51.7\%) \\
Negative & 243(60\%)                                  & 383(94.6\%)                                                       & 199(48.3\%)          \\\hline
\end{tabular}
\end{table*}

\begin{table*}[]
\centering
\caption{The statistics of using phone on crosswalks in Wuhan(empirical research, N=2901)\citep{RN389}}
\label{table 8}
\begin{tabular}{m{2cm}<{\centering}m{3cm}<{\centering}m{3cm}<{\centering}m{3cm}<{\centering}m{3cm}<{\centering}} \hline
Behaviors                  & \multicolumn{3}{c}{Use phones when crossing the road} & \begin{tabular}[c]{@{}c@{}}Not use phones \\when crossing the road\end{tabular} \\\hline
\multirow{2}{*}{Percentage} & Watch phone    & Have a call    & Listen to music    & \multirow{2}{*}{88.24\%}                                                          \\
                            & 6.65\%         & 2.83\%         & 2.28\%             &                                                                                  \\\hline
\end{tabular}
\end{table*}

\begin{figure*}[pos=!t]
\centering
\includegraphics[width=7in]{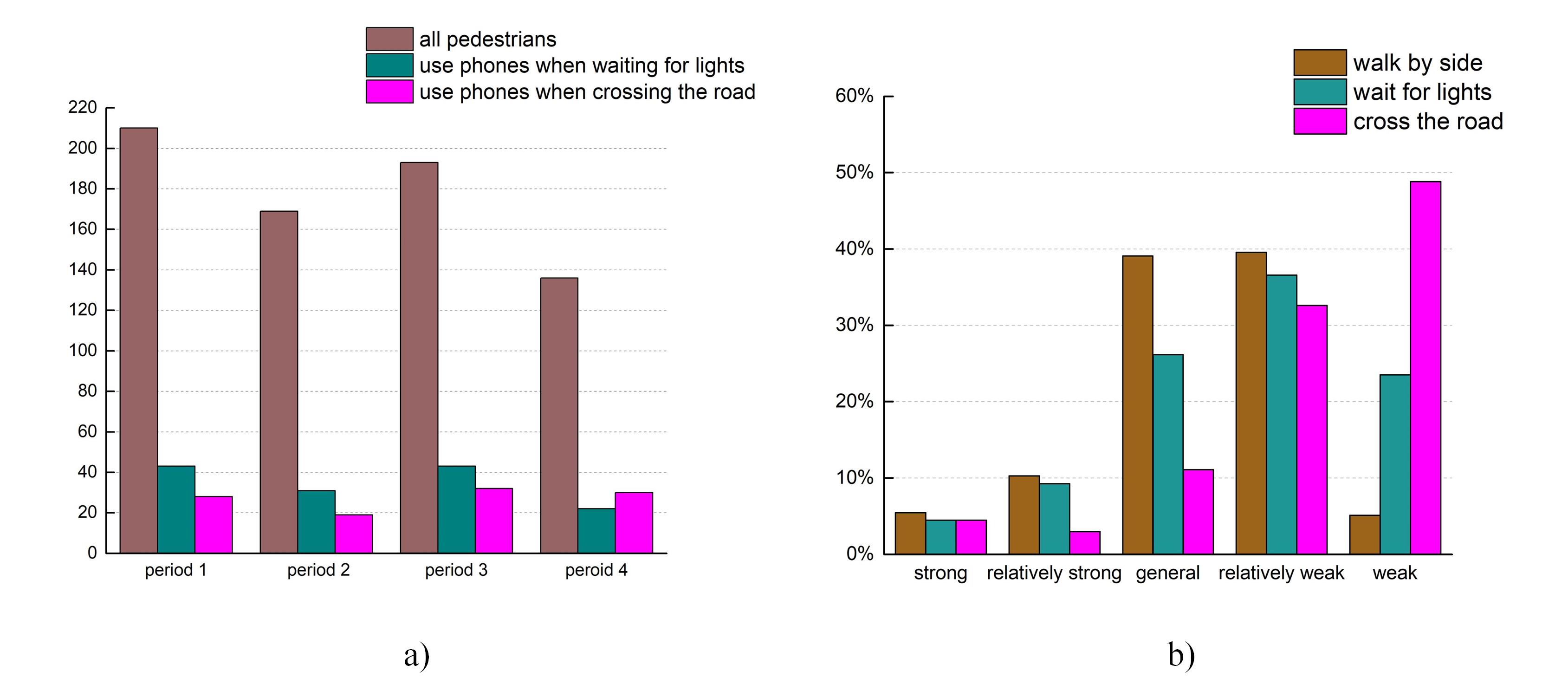}
\DeclareGraphicsExtensions.
\caption{The phone usage of pedestrians in different scenarios: a) Beijing (empirical research)\citep{RN462} b) Chongqing (questionnaire)\citep{RN463}}
\label{fig.6}
\end{figure*}

In America, a report of Liberty Mutual Insurance in 2013 surveyed 1,004 adults and found that 60\% of them had used phones when crossing the crosswalks\citep{RN274}. Moreover, the Consumer Product Safety Commission of America estimated that 3.76\% of American injuries are related to phone usage and Nasar et al. \citep{RN276} thought injuries caused by phone usage would be much bigger. In England, a report in 2019 found that 31.37\% of adolescent pedestrians crossed the crosswalks using phones\citep{RN277}. Similarly, phone usage of pedestrians has been an important problem of Chinese pedestrian safety.The table \ref{table 7} as well as \ref{table 8} are statistics of phone usage on crosswalks in Hefei and Wuhan respectively while Fig. \ref{fig.6} shows data collected in Beijing and Chongqing. From these data, it could be concluded that about 12\% of pedestrians would really utilize phones when crossing the road. What is more, about half of the pedestrians thought they should not be punished because of using phones\citep{RN388}. This reflects a lack of awareness to obey traffic rules of Chinese pedestrians.  As can be seen from table 8, watching telephone screens, having a call and listening to music are three aspects of phone usage and watching phones is the most common behavior. During these processes, the attention of pedestrians is occupied so they hardly perceive surroundings and communicate road rights with vehicles. \\
Ling et al.  emphasized that the time to cross a road would increase and the times pedestrians watch around would decrease when they use phones\citep{RN388}. According to a research\citep{RN286}, the proportion of pedestrians killed while using phones has increased by more than 3.5\% in 2010. Additionally, pedestrians using phones are more likely to conflict with vehicles than pedestrians not using phones. \\
There are two scenarios where pedestrians use their mobile phones at the crosswalk; the first is when the pedestrian traffic light is green and the second is when the pedestrian traffic light is red. In the first case, pedestrians might be relatively safe since pedestrians have the road right while cars have to wait. However, drivers might run red lights when pedestrians cross the road using phones. In this situation, distracted pedestrians cannot avoid a collision. In the second case, it will be extremely unsafe for pedestrians. Firstly, cars may not realize pedestrians are utilizing phones and think pedestrians would behave like normal persons. Furthermore, the distraction would hinder the intent communication between pedestrians and vehicles. As a result, drivers do not know how to respond to the behaviors accordingly and cause accidents. \\   
From the analyses above, we could conclude the unsafety comes from little communication and poor detection when pedestrians are utilizing phones, which highlights the importance of communication as well as detection again. In the context of traditional cars, drivers could partially recognize who are using phones and slow down to keep safe. Nevertheless, the intent communication remains a problem. Under the context of AVs, could AVs detect phone utilization and communicate with pedestrians well? 

\subsection{The summary of pedestrians' technical demands for AVs}
Based on the above phenomena and analyses, pedestrian safety requires good detection and communication between cars and pedestrians. In the following, the technical demands of pedestrians for AVs would be summarized. Simultaneously, we propose criteria to evaluate the ability of each technology. \\
In the detection part, in order to protect the safety of normal pedestrians (under no bad behaviors), vehicles should know the location and features of pedestrians. Moreover, occluded pedestrian detection is required to solve the problems of occlusion, which often takes place in red lights running. Additionally, remote(small) pedestrian detection is essential to protect pedestrians when pedestrians are jaywalking. Last but not the least, phone utilization detection is required to distinguish normal pedestrians and distracted ones. To evaluate the detection ability, the precision of location and classification will be reviewed and summarized. \\
In terms of communication, other than existing physical information transferring, more interfaces should be designed for interaction. For example, more external interfaces are needed to transfer the intent of cars to pedestrians in the scenarios of occlusion as well as jaywalking. As mentioned above, we suggested the state of AVs should be well transferred, which would be worked as our evaluating metric. To make it concrete, we propose the intent transferring comprised of stopping, accelerating, slowing down, as well as the observation of pedestrians, to evaluate the ability of communication.   \\
Importantly, there is a hidden factor, receptivity, that plays an important role in the relationship between AVs and pedestrians. If pedestrians cannot accept AVs, the communication is meaningless for pedestrians could not trust the information. Furthermore, there will be significant obstacles to bring AVs into the market. In this article, we have considered more about Chinese pedestrians, hence, we take the receptivity of Chinese pedestrians into account. \\
In summary, the adaptability of AVs depends on the adaptability of communication, detection as well as receptivity and hence we evaluate the adaptability of the three technologies to figure out the adaptability of AVs. The following section 3 will talk about the adaptability of AVs to pedestrians in detail. At the same time, the adaptability relationship would change the current mentality of pedestrians and ask for the government’s regulations adjustment. Based on the adaptability, we would simply analyze the influence that AVs would bring to current pedestrians mentality and government regulations.   

\section{The Adaptability of Driverless Technologies to Pedestrians}
China has a challenging driving environment of pedestrians, in which running red lights, jaywalking as well as phone usage take place usually. These behaviors lead to crowd, occlusion and poor communication, greatly harming the safety of pedestrians. Analyzing the scenarios, detection, communication and receptivity are summed up as key technical demands for safety. As a promising technology, AVs could perceive the world through multiple sensors, so can AVs solve these problems? What are the essential driverless technologies?\\
The adaptability could be divided into two parts: AVs adapt to pedestrians and pedestrians adapt to AVs. Corresponding to the demands, driverless technologies that are dedicated to solving these problems are divided into detection, interaction, and receptivity. In these researches, detection applies various sensors and algorithms to determine whether there are pedestrians and find their location, which shows the ability of AVs to feel the world and hence describes the adaptability of AVs to pedestrians. According to the demands, normal pedestrian detection, occluded pedestrian detection, small pedestrian detection and distracted pedestrian are reviewed and analyzed respectively. Furthermore, the technology of interaction utilizes different kinds of external interfaces to transfer the state of AVs to pedestrians, which shows the ability of pedestrians to understand AVs and hence describe the adaptability of pedestrians to AVs. And the researches of receptivity are committed to seeking factors that influence the acceptance of pedestrians to AVs; this reflects the pedestrians’ trust in AVs and also describes the adaptability of pedestrians to AVs.        \\

The following will examine these technologies by reviewing articles. After that, we would give out the adaptability by combining the analyse of phenomenon with technologies. To get an intuitive result, three levels of descriptions are applied to evaluate the adaptability: bad is used for poor research status and terrible adaptability; OK is used for good research status but there is still room to improve; excellent is used for splendid work and good adaptability.   \\
Furthermore, after summarizing the adaptability, the influence on these driverless technologies to the mentality of pedestrians and government regulations would be talked about in a simple manner, to offer some guidance to Practitioners.

\subsection{AVs adapt to pedestrians: detection}
As discussed above, detection is important to guarantee pedestrian safety. As a promising technology, AVs pay more attention to the detection task to recognize and localize pedestrians. However, the occlusion, small size as well as distraction of pedestrians make it a challenging task in China. To recognize the capabilities, the detection task would be divided into four subtasks including normal, occluded, small and distracted pedestrian detection.

In order to better review the ability, we first summarized the pedestrian dataset.        
\subsubsection{Pedestrian dataset}
     The popularity of data-based methods promotes the importance of datasets. The improvement of dataset also reflects the technical tendency of pedestrian detection. The authoritative pedestrian datasets are summarized in Table \ref{table 9}.  \\
\begin{table*}[]
\centering
\begin{threeparttable}
\caption{Comparison of pedestrian detection datasets}
\label{table 9}
\begin{tabular}{m{2.6cm}<{\centering}m{1cm}<{\centering}m{1.8cm}<{\centering}m{1.8cm}<{\centering}m{1.8cm}<{\centering}m{1cm}<{\centering}m{3cm}<{\centering}m{1cm}<{\centering}} \hline
Dataset                                                    & Pedestrian number       & \multicolumn{3}{c}{Occlusion labels}                                                                      & Evaluation Metrics\tnote{1}        & Best detector \& performances        & Year                        \\\hline
MIT\citep{RN362}                           & 924                      & \multicolumn{3}{c}{YES}                                                                                     & MR-FPPW                  & (HOG,100\%)                          & 2000                        \\
USC\citep{RN363}                           & 816                      & \multicolumn{3}{c}{NO}                                                                                     & ROC\tnote{2}                       & -                                    & 2005                        \\
INRIA\citep{RN370}                         & 1774                     & \multicolumn{3}{c}{NO}                                                                                     & MR-FPPW                  & (F-DNN, 7\%)                        & 2005                        \\
Daimler\citep{RN365}                       & 4000                     & \multicolumn{3}{c}{NO}                                                                                     & ROC                      & (MLS,28\%)                         & 2006                        \\
CVC\citep{RN364}                           & 1000                     & \multicolumn{3}{c}{NO}                                                                                     & ROC                      & -                                    & 2007                        \\
ETH\citep{RN367}                           & 12,000                   & \multicolumn{3}{c}{NO}                                                                                     & R-FPPI                   & (RPN+BF, 30\%)                      & 2007                        \\
TUD-Brussels\citep{RN368}                  & 3247                     & \multicolumn{3}{c}{NO}                                                                                     & P-R                      & (SpatialPooling,47\%)               & 2009                        \\
\multirowcell{2}{Caltech\\\citep{RN366}\\\citep{ RN372}} & \multirow{2}{*}{350,000} & \multicolumn{3}{c}{YES}                                                                                     & \multirow{2}{*}{MR-FPPI} & \multirow{2}{*}{(AR\-Ped,6\%)}       & \multirow{2}{*}{2009}       \\\cline{3-5} 
                                                           &                          & \begin{tabular}[c]{@{}c@{}}No\\   occlusion(0\%)\end{tabular} & Partial (1\%-35\%)  & Heavy                    (35\% - 80\%) &                          &                                      &                             \\
\multirowcell{2}{KITTI}\\\\\citep{RN371}        & \multirow{2}{*}{80,000}  & \multicolumn{3}{c}{YES}                                                                                     & \multirow{2}{*}{P-R}     & \multirow{2}{*}{(FichaDL, 81.73\%)} & \multirow{2}{*}{2012(2015)} \\\cline{3-5}
                                                           &                          & Easy(0-15\%)                                                 & Moderate     (15\%-30\%) & Hard(35\%-50\%)     &                          &                                      &                             \\\hline
\end{tabular}
   \begin{tablenotes}
     \item[1] MR, FPPW, FPPI, R and P respectively infer to miss rate, false positive per window, false positive per image, recall and precision. the higher, the better in MR-FPPW
and P-R curve whereas the lower, the better in ROC,  MR-FPPI, MR-FPPW and R-FPPI curve. 
     \item[2] ROC curve takes true positive rate and false negative rate as coordinate
  \end{tablenotes}
\end{threeparttable}
\end{table*}

These datasets are mainly used to evaluate the capabilities of detectors, of which Caltech dataset as well as KITTI dataset are popular datasets in recent years. The number of pedestrians is related to the number of ground truth and frames, representing the difficulty of detection. Caltech dataset ranks first with 350,000 pedestrians at 1,000,000 frames. Additionally, the dataset released later often has a larger number of pedestrians because data-based methods dominate the mainstream. As the year increases, occluded labels are added to the dataset more because occlusion usually occurs in real scenarios, which is why KITTI as well as Caltech datasets become popular currently. The occlusion level could be evaluated by the fraction of occlusion, which can be calculated as one minus the fraction of visible area over the total area. However, KITTI and Caltech datasets have different occlusion labels. KITTI dataset divided scenes into three levels: easy, moderate and heavy whereas Caltech grouped them into no, partial occlusion as well as heavy occlusion based on occlusion ratio, which is shown in the third column. Naturally, the two datasets both take the occlusion into account when evaluating detectors.\\

To emphasize different things, the metrics utilized to evaluate the abilities of detectors are different. The ROC curve evaluates the detectors in terms of true positive rate and false negative rate in the early dataset. The MR-FPPI curve used by Caltech and ETH reflects the relationship between miss rate (MR) and false positive per image (FPPI), and used miss rate value when the FPPI value is 10-1   as the evaluation metric. In order to emphasize precision and recall, KITTI dataset used the P-R curve to evaluate the performance of the detectors and utilize the area between the P-R curve and coordinates as the evaluation metric. Recently, the P-R and MR-PFFI curves are utilized more in the detection task due to the popularity of KITTI and Caltech datasets.  Reasonably, each dataset has an optimal detector shown in the fifth column. We cannot get a specific optimal detector in some datasets so we use a dash to represent. In the early dataset, detectors gain a splendid detection rate even around 100\% while detectors get worse in recent as a result of increasing difficulty in datasets.  Additionally, the popularity of neural network-based methods are reflected in terms of the name of optimal detectors. More importantly, the websites of these datasets would rank detection methods to help researchers follow the latest development and this would be our approach to evaluate the capabilities. \\
    Last but not the least, KITTI dataset collected pictures from campus, city, road, residential and person scenes, which is closer to real scenarios in an autonomous driving environment. However, the urban Chinese pedestrians would be only talked in this paper so Caltech dataset would be more suitable. \\

\subsubsection{Normal pedestrians detection}
\begin{figure*}[pos=!t]
\centering
\includegraphics[width=7in]{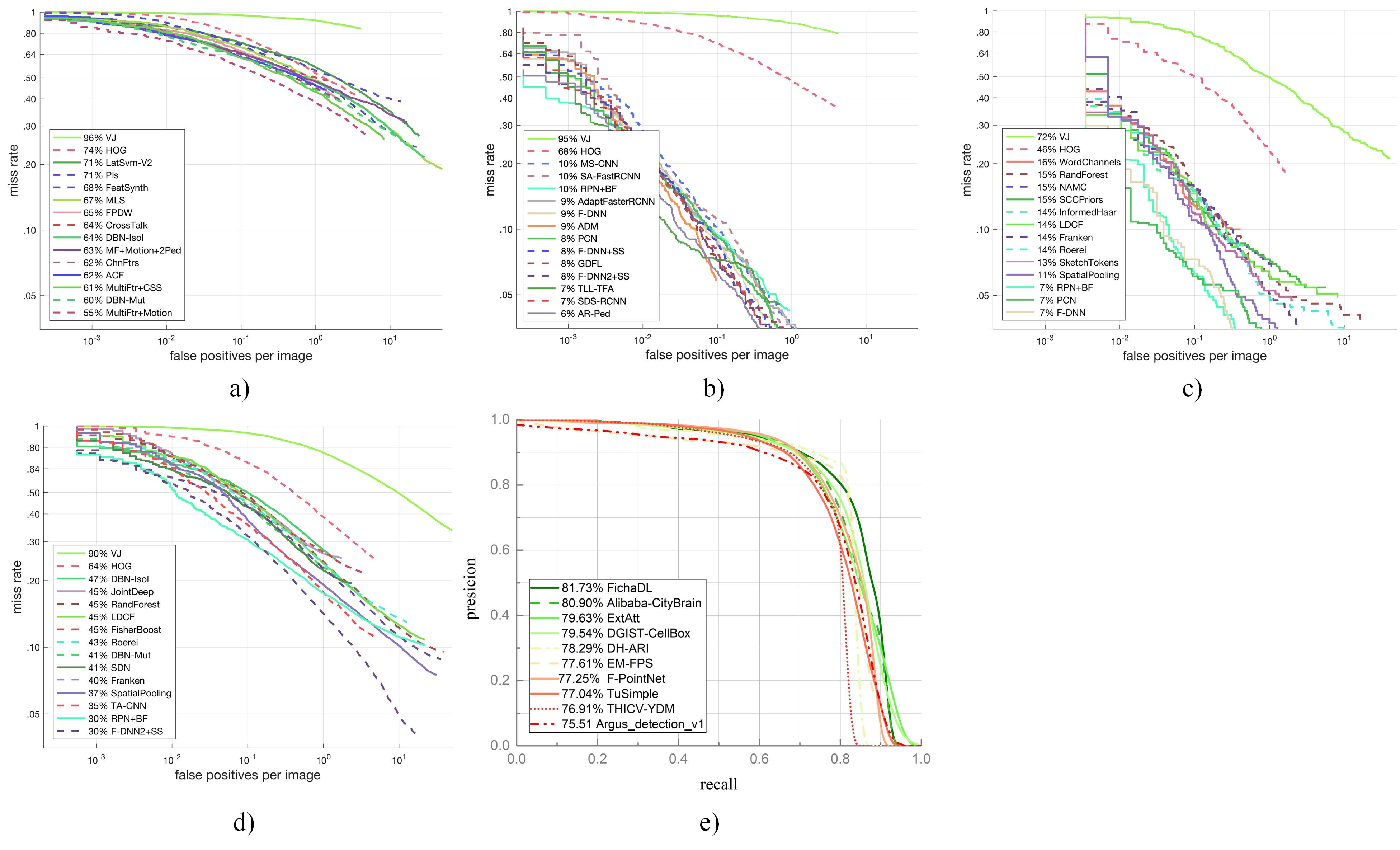}
\DeclareGraphicsExtensions.
\caption{The performances of 10 top detectors in five popular datasets: a) Caltech training \citep{RN366, RN372} b) Caltech testing\citep{RN366, RN372} c) ETH \citep{RN367}  d) INRIA \citep{RN370} e) KITTI \citep{RN371}(Data is summarized in October 1, 2019)}
\label{fig.7}
\end{figure*}

       Normal pedestrians are pedestrians who do not perform bad behaviors. We set the occlusion level, size as well as the distraction of these pedestrians to normal values. Normal pedestrian detection could assess basic detection capabilities because normal pedestrians are the largest parts of pedestrians. To check the ability, the performance of detectors is summed up based on four popular pedestrian datasets in Fig. \ref{fig.7}. The moderate and reasonable occluded levels of KITTI and Caltech datasets are selected to meet the demands of normal pedestrians.      \\
         It is observed that detectors in Caltech are the best with the lowest miss rate of 6\% among all the four datasets. The best detector in INRIA is slightly behind with a 7\% precision. However, the optimal detector named F-DNN2+SS in ETH dataset achieved the best miss rate of just 30\%. This is possibly because it is so old-fashioned that the latest detectors would not validate performance on it. Interestingly, the difference between Caltech training dataset and Caltech testing dataset is huge, which is because data-based methods use a testing dataset to validate while features-based methods use a training dataset. This also reflects the superiority of data-based methods.\\
In KITTI dataset, the area of the curve is utilized to evaluate detection performance and the best detector achieves an average moderate precision of 81.73\%. Roughly speaking, we could say the best detectors in Caltech with 7\% miss rate is better than in the best detectors in KITTI with 81.73\% precision. As talked above, the scenarios in Caltech dataset are more suitable for the setting of Chinese urban pedestrians, hence, the result of Caltech is more reliable in this article. In conclusion, the capabilities of normal pedestrian detection are excellent with a miss rate of 6\% and the driverless technology of normal pedestrian detection is adaptive to urban China.\\

\subsubsection{Occluded pedestrians detection}
Pedestrian occlusion usually happens in driving scenarios and about 70\% of pedestrians have occlusions according to Caltech dataset\citep{RN366, RN372}, which is among the toughest problems in pedestrian detection. Since Chinese pedestrians perform badly on crosswalks, occlusion has always been a serious problem. Compared to no occlusion, occluded pedestrian needs two bounding boxes to jointly label, one for visible portion and one for full extent. Furthermore, occlusion could be divided into two categories: pedestrians are occluded by other objects, which often causes information missing and leads to false negatives, and pedestrians are occluded by other pedestrians, which brings lots of interfering information and leads to false positives\citep{RN373}. \\
 To figure out the capabilities, the top 10 detectors in KITTI and Caltech datasets are summarized in terms of different occlusion levels in Table \ref{table 10}. The hard level of KITTI dataset and the heavy level of Caltech dataset are the most occluded scenarios.

\begin{table*}[]
\centering
\begin{threeparttable}
\caption{The performances of 10 top detectors of KITTI \& Caltech datasets in different occlusion levels(data is summarized in October 1, 2019)}
\label{table 10}
\begin{tabular}{m{2.5cm}<{\centering}m{1cm}<{\centering}m{2.5cm}<{\centering}m{2cm}<{\centering}m{2.5cm}<{\centering}m{1cm}<{\centering}m{2cm}<{\centering}m{2cm}<{\centering}} \hline 
\multirow{2}{*}{Rangking \& Methods}          & \multicolumn{3}{c}{KITTI dataset\tnote{1}} & \multirow{2}{*}{Rangking \& Methods}      & \multicolumn{3}{c}{Caltech dataset\tnote{2}} \\
                                              & Easy          & Moderate/Change              & Hard/Change\tnote{3}                 &                                           & No               & Patial/Change                 & Heavy/Change                 \\\hline
1.     FichaDL                                & 88.27\%       & 81.73\% /6.54\%       & 75.29\%/12.98\%      & 1. AR-Ped\citep{RN286}    & 5.00\%           & 12.00\%/7\%            & 49.00\%/44\%          \\
2.     Alibaba-CityBrain                      & 88.13\%       & 80.90\%/7.23\%        & 74.08\%/14.05\%      & 2. SDS-RCNN\citep{RN391}  & 6.00\%           & 15.00\%/9\%            & 59.00\%/53\%          \\
3.     ExtAtt                                 & 87.95\%       & 79.63\%/8.32\%        & 74.78\%/13.17\%      & 3. F-DNN+SS\citep{RN392}  & 7.00\%           & 15.00\%/8\%            & 54.00\%/47\%          \\
4.     DGIST-CellBox                          & 87.77\%       & 79.54\%/8.23\%        & 75.70\%/12.07\%      & 4. F-DNN\citep{RN392}     & 7.00\%           & 15.00\%/8\%            & 55.00\%/48\%          \\
5.     DH-ARI                                 & 87.43\%       & 78.29\%/8.32\%        & 69.91\%/17.52\%      & 5. PCN\citep{RN394}       & 7.00\%           & 16.00\%/9\%            & 56.00\%/49\%          \\
6. EM-FPS                                     & 84.93\%       & 77.61\%9.14\%         & 72.52\%/12.41\%      & 6. F-DNN2+SS\citep{RN393} & 6.00\%           & 16.00\%/10\%           & 40.00\%/34\%          \\
7. F-PointNet\citep{RN374}    & 87.81\%       & 77.25\%/7.32\%        & 74.46\%/13.35\%      & 7. GDFL\citep{RN395}      & 6.00\%           & 17.00\%/11\%           & 43.00\%/37\%          \\
8. TuSimple\citep{RN376, RN375} & 86.78\%       & 77.04\%/10.56\%       & 72.40\%/14.38\%      & 8. ADM\citep{RN397}       & 7.00\%           & 18.00\%/11\%           & 30.00\%/23\%          \\
9. THICV-YDM                                  & 87.27\%       & 76.91\%/9.74\%        & 69.02\%/18.25\%      & 9. TLL-TFA\citep{RN398}   & 6.00\%           & 18.00\%/12\%           & 29.00\%/23\%          \\
10. Argus-detection-v1                        & 83.49\%       & 75.51\%/10.36\%       & 71.24\%/12.25\%      & 10. MS-CNN\citep{RN396}   & 8.00\%           & 19.00\%/11\%           & 60.00\%/52\%   \\\hline       
\end{tabular}
   \begin{tablenotes}
     \item[1]  KITTI dataset uses P-R curve to evaluate detectors and the higher, the better.
     \item[2]  Caltech dataset uses MR-FPPI curve to evaluate detecors and the lower, the better.
     \item[3]  Change denotes the value of current level minus the value of first column of each dataset;
  \end{tablenotes}
\end{threeparttable}
\end{table*}

It is reasonable that when occlusion levels improve, accuracies would significantly decrease. The detectors in easy occlusion of KITTI has an excellent performance since 88.27\% of pedestrians are detected (the same is true in no occlusion in Caltech). However, the performance becomes particularly poor, especially in heavy occlusion in Caltech (hard in KITTI), for 49\% of pedestrians are missed. More importantly, the change rate is huge when the difficulty of scenarios changes from easy to hard, especially in Caltech dataset with a change rate of about 40\%. This is because Caltech dataset has a more complicated occlusion status. However, occlusion is common and sometimes there is heavy occlusion taking place in jaywalking or crosswalk crossing scenarios. Therefore, both the easy level and heavy level of occlusion detection are important in the Chinese urban environment. Above all, driverless technologies perform excellent detection in easy occlusion whereas detectors perform very poorly in the heavy level. Considering the importance of easy occlusion and heavy occlusion, we think driverless technology has overall bad research in occluded pedestrian detection due to the unacceptable result of heavy occlusion.

\begin{figure*}[pos=!t]
\centering
\includegraphics[width=7in]{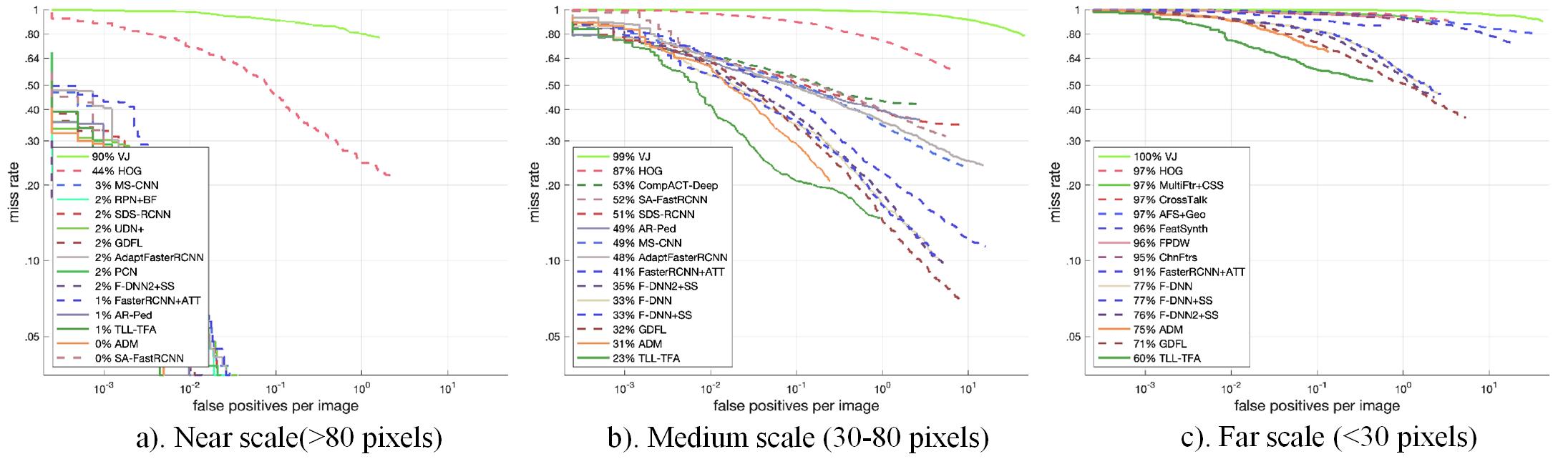}
\DeclareGraphicsExtensions.
\caption{Small pedestrian detection in three scales in Caltech dataset(Data is summarized in October 1, 2019)}
\label{fig.8}
\end{figure*}

\subsubsection{Small pedestrians detection}
Small pedestrians mean that pedestrians take up small parts of pixels in the eyes of AVs. As analyzed above, jaywalking is very common and has caused a number of accidents in the Chinese environment. In the scenarios, the speed of vehicles is fast and they need to detect pedestrians at long distances to take timely measures. Therefore, small pedestrian detection is important to protect pedestrians. However, small pedestrian detection is complicated due to low resolution and noisy presentation\citep{RN382}.\\
In the authoritative dataset, COCO of Microsoft\citep{RN381}, they define objects smaller than 32*32 pixels as small objects. In the KITTI dataset, researchers set a minimized bounding box height of 25 pixels for moderate as well as hard level while Caltech dataset categorized them into near scale, medium scale and far scale. Additionally, Dollar et al. \citep{RN366}thought that pedestrians from 30 to 80 pixels are most important for automotive settings, respectively correspond to 20m and 60m away from the pedestrians at urban speed of 15m/s. Above all, we would focus more on pedestrians with the range from 30 to 80 pixels, which include small pedestrians in autonomous settings.   \\
In the COCO dataset, general small object detection reflects an average precision of 0.343, compared to medium objects of 0.556 and large objects of 0.660. This also shows the difficulty of small object detection. As for small pedestrian detection, Caltech researchers summed up the performance of detectors in three scales shown in Fig. \ref{fig.8}. Reasonably, detectors in near scale has the best performance with the lowest miss rate of nearly 0\%. However, when it comes to medium scale, the performance drops dramatically with the lowest miss rate of 23\%. To make matters worse, detectors in the far scale have a terrible miss rate of only 60\%. As analyzed above, the medium-scale (30-80 pixels) pedestrian data is more suitable for Chinese urban pedestrians. Therefore, the best performance is the TLL-TFA algorithm with a miss rate of 23\%. In summary, driverless technology is OK to detect small pedestrians but has some room to improve. \\

\subsubsection{Distracted pedestrians detection}
Distraction usually happens when pedestrians are reading books, talking to others as well as using phones and phone utilization is the main reason. Distraction can seriously harm the safety of pedestrians as discussed above. In the context of AVs, it is essential to detect pedestrians using phones and accordingly decide how to respond. \\
Although the phone utilization of drivers gains much attention in research, there is a lack of researches about pedestrians using phones. Akshay et al.  proposed a vision-based framework to classify whether a pedestrian is using a phone, with the highest accuracy of 91.20\%\citep{RN383}. At the same time, a pedestrian dataset using phones was proposed as well, in which pedestrians are in high resolution. However, there is little research to further the study. \\
Yet, surprisingly, phone utilization hasn’t always been terrible for AVs. Vehicle-to-pedestrians(V2P) communication is thought as a method to guarantee active safety, in which smartphone works as a receiving device. Ahmed et al.\citep{RN127} proposed a V2P-based application to broadcast alert information, including possible collision distance and time, to both phone users and AVs. At the same time, Pooya et al.\citep{RN342} proposed a model using phones to send alert warnings to pedestrians when they are around the crosswalk. Furthermore, He et al.\citep{RN343} proposed a V2P model utilizing WIFI, Bluetooth as well as DSRC technology to establish interaction between AVs and pedestrians.  In these cases, the smartphone plays an important role to avoid collision rather than lead to accidents.\\

All in all, there exists research with excellent phone utilization detection of 91.20\%, nevertheless, no other researches are conducted to support and further the study. Moreover, the smartphone has played a totally new role to interact with the pedestrians rather than ruin the safety, and the method performs well. In conclusion, there is not systematic work to support the accurate detection to distracted pedestrians but the phone has become another way to protect safety.  Hence, we think the driverless technology is OK and has some room to promote. 
\subsubsection{The adaptability of detection}
To summarize the above conclusions, normal pedestrian detection is excellent. Occluded pedestrian detection is bad. Distracted pedestrian detection is OK. Small pedestrian detection is OK.
\subsection{Pedestrians adapt to AVs: interaction}
According to Gunnar\citep{RN280}, human-machine interaction(HMI) transfers communication information between human users and machines via human-machine interfaces. Under the context of traffic, the interaction takes place between vehicles and road users, such as pedestrians, and human-machine interfaces are utilized as the interacting methods. When pedestrians enter the road network, they begin constant information exchange with the traffic environment \citep{RN68}. Traditionally, drivers interact with pedestrians to negotiate road rights. While under the context of AVs, drivers do not have the control right of the cars (above Level 4) and cannot know the state. Sometimes the person sitting on the seat might be distracted, leaving pedestrians to infer the state of AVs alone. In this case, pedestrians could get limited information by observing the speed and distance. There is a special need for alternative communication techniques of AVs, which must be able to substitute for the gaze of the driver\citep{RN157}. Furthermore, the interaction helps to boost acceptance and develop proper mental models when AVs are originally put into markets. More importantly, it is crucial to inform pedestrians when AVs have a failure state\citep{RN134}.\\ 
\subsubsection{Human Machine Interfaces}
Recently, a lot of researches has been conducted on interaction, focusing on the human-machine interfaces. The used interfaces and transferring information are summarized in Table \ref{table 11}. Google developed a patent utilizing LED displays to notify pedestrians of the state of AVs\citep{RN451}. Tobias et al. \citep{RN135} used LED strips to conduct interaction, in which lights are laid in a line and different locations as well as light numbers show different information. The middle four lights light up to express autonomous mode is on; lights expand to the sides, indicating AVs notice pedestrians while the lights shrink from the sides to the middle, which means AVs is about to start; all lights light up to indicate AVs are resting. \\

 Karthik et al. proposed a fusion interface to communicate with pedestrians. In their studies, infrastructures and electric devices can also become important interfaces of interaction\citep{RN65}. They have proposed four prototypes and prototype 1 utilized a speaker coupled with LED strips on cars. In order to transfer intent, the LED strip was mounted on the vehicle and exhibit four states. Solid red lights indicate the pedestrian shouldn’t cross as the vehicle would not stop. Blinking blue lights mean the vehicle was aware of the pedestrian. Green lights moving from left to right indicate the vehicle had fully halted and it was safe to cross. Purple lights moving from right to the left meat the vehicle would start soon. \\

\begin{figure*}[pos=!t]
\centering
\includegraphics[width=7in]{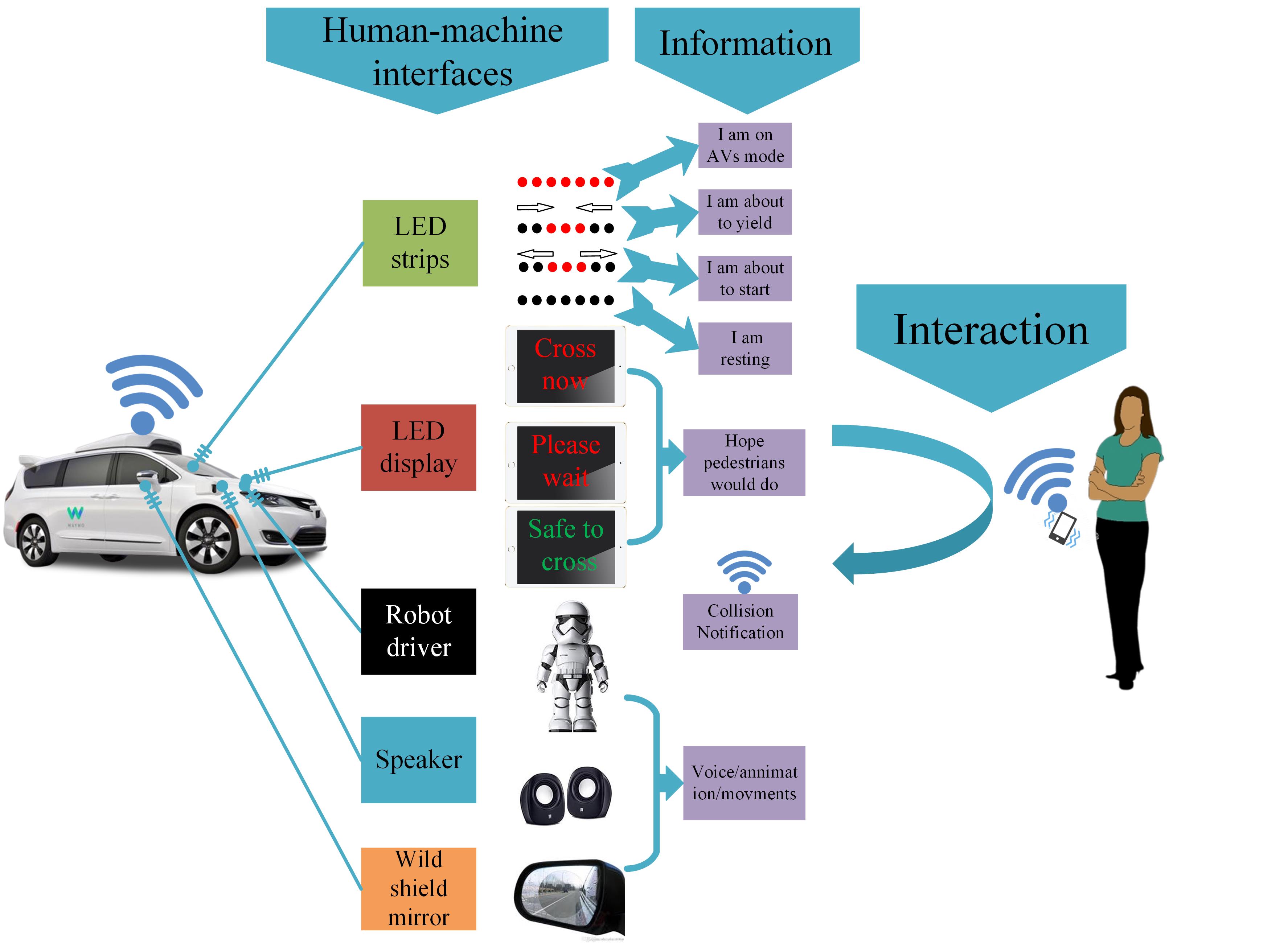}
\DeclareGraphicsExtensions.
\caption{External interfaces and transferring information for interaction in driverless technology}
\label{fig.9}
\end{figure*}

Milecia et al. proposed a method fusing the speaker, LED word displays as well as strobe lights to transfer intent and corresponding experiments showed pedestrians’ trust to AVs was greatly improved\citep{RN132}. Furthermore, based on human-robot interaction, three novel methods were proposed by Nicole et al\citep{RN71}. First, Gaze and gestures of conventional drivers are logged and projected on wild shield screens to interact with pedestrians. Second, elements in the front of cars, such as headlights, radiator grill as well as side mirrors, were also utilized to make gestures like drivers to inform pedestrians. Third, a robot driver was utilized to imitate the eye contact and gestures of drivers.  \\
        In experiments of Rahimian et al.\citep{RN342}, pedestrians could receive an alarm auditory from the AVs when they are crossing the road, alerting not using phones and paying attention to driving environment. However, phones are not always terrible during the interaction process. Ahmed et al.\citep{RN127} applied phones to interact with pedestrians. Based on GPS data, phone application and AVs could calculate the distance and time to the collision point as well as danger indexes, therefore, the application could warn pedestrians and AVs by displaying a warning message or vibrating. Additionally, He et al.\citep{RN343} proposed a V2P model which utilized WIFI, Bluetooth as well as DSRC communication technology to build interaction between AVs and pedestrians.  
       Last but not the least, it will be essential to transfer information to pedestrians under the collapse of AVs, so Aaron et al.\citep{RN134} utilized the words of iPad and LED strips to send faulty information to pedestrians.  \\
      Through these external interfaces, pedestrians argued they know more about the state of AVs in \citep{RN68, RN69, RN78, RN361, RN65, RN132, RN360} and thereafter they trust AVs more in \citep{RN357, RN74, RN68, RN69, RN127, RN78, RN132}. To show the process of interaction, the features and transferring information are partially plotted in Fig. \ref{fig.9}.  \\
\subsubsection{Shortages and suggestions on HMI design}
Nevertheless, every interface has its inevitable shortages. Visual interfaces are the most commonly used interfaces but it is not useful for people with color blindness, visual impairment and distraction. Auditory interfaces are thought to be more like a command and hence pedestrians dislike it. In addition, pedestrians would be confused when there are lots of AVs and other sounds. Phone vibration isn’t preferred by pedestrians for there are other functions that would cause vibration in a phone \citep{RN65}. \\
To tackle the shortages, a fusion of multiple features is a promising method to transfer more information and reach higher robustness\citep{RN65}. In Karthik’s experiments, mixed interfaces got the best score. However, it doesn’t mean the more interfaces mixed, the better interaction it gets. However, information overload takes place when there are too many interfaces\citep{RN65}.  In the situations, pedestrians tend to check all interafces and then allow a go-head, thus leading to inefficiency. Therefore, researchers should consider possible information overload when fusing different interfaces. Furthermore, interaction needs a standard language to decide what interfaces to use and how to use.  At above, different interfaces and different infomration transferrring methods are adopted by various researchers. Considering the future with AVs, pedestrians would get confused when interacting with AVs of different brands\citep{RN133}.  Therefore, a standard interaction language is extremely needed for better welcoming AVs.\\

\subsubsection{The adaptability of pedestrians-AVs interaction}
In conclusion, equipped with external features, AVs could transfer the following state: about to start, about to stop, slowing down, starting AV mode as well as observing the pedestrians, etc. Furthermore, phone utilization would be warned and pedestrians could get the failure state of AVs.  Through these, pedestrians could know the driving state of AVs and hence decide what and how to do. Compared to traditional interaction methods, the external features of AVs seem to be more abundant and intuitive. Yet, there still exits some shortages in these researches and more efforts should be taken in the area of making a uniform interacting language and fusing multiple external interfaces. Above all, we conclude driverless technology is OK on interacting with pedestrians but still has room to improve. Moreover, we boldly predict that AVs could even reach a better interaction compared to traditional cars by accomplishing the current limitations. 
\subsection{Pedestrians adapt to AVs: receptivity}
Receptivity was originally defined as the willingness to accept uncertain, unfamiliar, or paradoxical idea\citep{RN384}. Therefore, the receptivity of pedestrians to AVs is the willingness of pedestrians to accept AVs, similar to the concept of trust and acceptance.  \\

The high receptivity of pedestrians is important. On the one hand, as Vahidi concluded, the main obstacle to achieve a place in the market is not only technical issues but also the lack of acceptance of new ideas, which is important to gradually push AVs into markets\citep{RN291}. On the other hand, pedestrians would be more willing to interact with AVs and push the technology to evolve.  \
\subsubsection{Factors that influence receptivity}
Recent researches have paid attention to factors that influence receptivity, which is summarized in Table \ref{table 12}. Demography (including age \& gender), reflects some basic properties of persons and is thought to influence the receptivity. Shuchisnigdha found males and younger people tend to trust AVs due to more interest in new technologies\citep{RN74}.  However, Reig thought demographic variables were not meaningfully related to beliefs and perceptions\citep{RN40}. \\
Moreover, understanding to AVs has a strong relationship with receptivity.  Samantha et al.\citep{RN40} found an insufficient understanding of AVs would lead to mistrust and made-up explanations for the behaviors of AVs. Monika et al.\citep{RN287} showed that people’s understanding of driverless algorithms improves receptivity and pedestrians now know a little how AVs conduct their movements

\clearpage
\onecolumn
\begin{landscape}
\begin{longtable}{m{1.5cm}<{\centering}m{2cm}<{\centering}m{1.5cm}<{\centering}m{4cm}<{\centering}m{4cm}<{\centering}m{6cm}<{\centering}}
\caption{Interacting methods of driverless technologies to communicate with pedestrians}
\label{table 11}\\\hline

Reference                                                                                & Experimental type                                                                                       & Date & Interfaces                                                                                                                                                                                                                                        & Information                                                                                                                                                                          & Conclusion                                                                                                                                                                                                                                                   \\\hline
Karthik et al.\citep{RN65}                                               & Empirical experiment \& questionnaire                     & 2018 & 1. Speaker + LED strip 2.     Speaker + LED lights (in street)  3.     Animated face + phone haptic(pedestrians) 4.   Printed hand + phone audio + LED lights (street) & (AVs) about to start; about to stop; fully stop; notice the pedestrians;
& (1) Interfaces help pedestrians attempt to cross; (2) Interfaces can exist in the environment; (3) AVs should use a combination of visual,auditory, and physical interfaces.                                   \\

Milecia et al.\citep{RN132}                                              & Empirical experiment \& simulating experiments          & 2017 &LED word display + speakers +strobe lights                                                                                                                                                         & (Pedestrians): cross now; stop; wait to cross                                                                                            & (1)  Humans react positively and more predictably when the intent of the vehicle is communicated (2) Pedestrians trust AVs more when there are interfaces or they have prior knowledge to AVs \\

Aaron et al.\citep{RN134}                                                & Empirical experiment \& simulating experiments \& interview & 2018 & LED strobe lights+ iPad display                                                                                                                                                                                                                       & (Pedestrians): please wait; safe to cross & (1) There exists possible confusion of interfaces (2) Pedestrians want to know the state of AVs rather than what they should do                                           \\

Nicole et al.\citep{RN71}                                                & N/A                                                                                                      & 2017 & 1.     Windshield screen 2. Head lights, radiator grill, the side mirrors 3. Robot driver  & Human-like gaze and gestures                                                                                                                                                         & N/A                                                                                                                                                                                                                                                           \\

Tobias et al. \citep{RN135}                                              & A Wizard of Oz approach                                                                                 & 2015 & LED light strip                                                                                                                                                                                                                                 &In AV mode; about to yield; about to start;  is resting &Interfaces can promote the interaction between pedestrians and vehicles.                                                 \\

Ahmed  et al. \citep{RN127}  & User study \& simulating experiments                                                                    & 2016 & Smartphones                                                                                                                                                                                                                                     & Time to collision; velocity; distance to collision etc.& External interfaces create good performance high detection rate and user satisfaction \\                                                                                                                            

Urmson  et  al. \citep{RN451} & N/A (patent)                                                                                             & 2015 & An electronic sign or lights, a speaker\                                                                                                                                          & What AVs  are doing or going to do                                                                             & The interfaces could replace the interaction between pedestrians and drivers               \\           
                                                                                                            
Pooya et al.\citep{RN342} & Simulating experiments                                                                                  & 2018 & Smartphones                                                                                                                                                                                                                                     &Warn pedestrians that "Take care around and there are AVs"                                                                                              & Informing  pedestrians by phones could improve the safety                                                                                                                                                  \\

He et al. \citep{RN343}    & Simulating experiment \& case study                                                                     & 2016 & Smartphones                                                                                                                                                                                                                                     &The location of AVs and pedestrians                                                                                             & Bluetooth technology combined with DSRC could be workable for active pedestrian protection.\\

Evelyn et al. \citep{RN69}                                               & Empirical experiment                                                                                    & 2016 & Speaker + LED display                                                                                                                                                                                                                           & 1. Whether pedestrians’ presence has been perceived by the autonomous vehicle  2. Broadcast the detection result of  LiDAR & The audio  cues and LED strips are useful for interaction and promote trust.                                                                                                                                    \\

Shuchisnigdha et al. \citep{RN68}                                        & VR experiment                                                                                           & 2018 & Speaker + LED display                                                                                                                                                                                                                           & 1. AVs are braking  2. Pedestrians are safe to cross & (1). Familiar sign for pedestrian, clear text and clear verbal message is preferred (2). The inclusion interfaces increase pedestrians’  receptivity                                                 \\

Azra et al. \citep{RN133}                                                & N/A                                                                                                      & 2018 & LED light strips                                                                                                                                                                                                                                 &In AV or manual mode; about to start; about to  yield &External  interfaces help to add up trust to AVs; Features call for standardization;          \\\hline

\end{longtable}
\end{landscape}
\clearpage
\twocolumn 

 agreeing with previous research of Rogers\citep{RN385}. Furthermore, Monika et al. emphasized the usability as  well as trialability would take pedestrians closer to AVs. The experiments of Samantha et al.\citep{RN40}found the more interest pedestrians have, the higher level pedestrians would trust AVs. Actually, usability, trialability and interest improve the receptivity by promoting the understanding of pedestrians.\\
The interaction between AVs and pedestrians also promotes the receptivity\citep{RN133}. In the context of AVs, interaction is conducted based on external interfaces. Samantha et al. \citep{RN40}emphasized the importance of external interfaces to promote receptivity due to transferring the intent as well as the current state of AVs. Saleh et al. highlighted the intent communication and called it the most critical sign to win the trust\citep{RN289}. Ahmed et al.\citep{RN127}adopted a V2P application to warn persons and found a promotion in trust toward AVs. Azrac et al. discovered that after interacting with AVs once, pedestrians could gain a close understanding of AVs, hence promoting trust and acceptance. Similarly, pedestrians trust AVs more in \citep{RN68, RN69, RN132} after interacting with external interfaces. \\
Interestingly, the brands of companies influence the receptivity to AVs. Samantha et al.\citep{RN40} discovered pedestrians trust the AVs of Uber because of its largeness, engineering quality while others doubt Uber for its fast speed to develop AVs . \\
Accidents related to AVs harm the receptivity because negative events are more visible compared to positive events\citep{RN386}. Small events, even totally no responsibility from AVs, could set off a number of resistances\citep{RN387}. Actually, most of the accidents should not blame AVs as Favarò et al. showed in \citep{RN292} that contributing factors come more from other the manoeuvers of other vehicles. \\

\begin{table*}[]
\centering
\caption{Influencing factors of pedestrians’ receptivity to AVs}
\label{table 12}
\begin{tabular}{m{6cm}<{\centering}m{2cm}<{\centering}m{2cm}<{\centering}m{2cm}<{\centering}m{2cm}<{\centering}}\hline
Title                                                                                                                 & Reference                     & Method                    & Year & Factors                                    \\\hline
A Field Study of Pedestrians and Autonomous Vehicles                                                                  & \citep{RN40}  & Questionnaire             & 2018 & Understanding \& Ability \& Brand \&Demography \\
Applied artificial intelligence and trust—The case of autonomous vehicles and medical assistance devices               & \citep{RN287} & Case study                & 2016 & Understanding                                  \\
Trust in AV: An Uncertainty Reduction Model of AV-Pedestrian Interactions                                             & \citep{RN78}  & VR                        & 2018 & Ability                                    \\
Towards Trusted Autonomous Vehicles from Vulnerable Road Users Perspective                                            & \citep{RN183} & Model (trust model)       & 2017 & Ability                                    \\
Development and validation of a questionnaire to assess pedestrian receptivity toward fully autonomous vehicles       & \citep{RN74}  & Questionnaire             & 2017 & Interaction \& Demography                  \\
P2V and V2P Communication for Pedestrian Warning on the basis of Autonomous Vehicles                                  & \citep{RN127} & Model (application)       & 2016 & Interaction                                \\
External Vehicle Interfaces for Communication with Other Road Users?                                                  & \citep{RN133} & Model (external interfaces) & 2018 & Interaction                                                              \\
Intent Communication between Autonomous Vehicles and Pedestrians                                                      & \citep{RN132} & Model (external interfaces) & 2017 & Interaction                                \\
Pedestrian Notification Methods in Autonomous Vehicles for Multi-Class Mobility-on-Demand Service                     & \citep{RN69}  & Model (external interfaces) & 2018 & Interaction                                \\
Investigating pedestrian suggestions for external interfaces on fully autonomous vehicles: A virtual reality experiment & \citep{RN68}  & VR                        & 2018 & Interaction                               \\\hline
\end{tabular}
\end{table*}

\subsubsection{The adaptability of receptivity}
Above all, demography, understanding, the ability of AVs, interaction, the brands of companies and accidents would influence the receptivity of AVs.  By analyzing these factors, interaction has taken up largely in the influencing factors, which emphasize the importance of interaction and interacting experiences again. Additionally, the understanding and the ability also contribute to the receptivity. More importantly, the ability of AVs, knowledge and interaction could promote each other and the commonality of these factors is to stress the real experiences with AVs. It is known that real experiences occur in close contact with pedestrians. However, we seldom see researches on Chinese pedestrians; we hardly see interacting experiment is carried out in China; we never see waymo (google) or cruise (general motor) test their AVs in the Chinese environment. Therefore, the receptivity of Chinese pedestrians to AVs must be very low.  In conclusion, the receptivity of Chinese pedestrians is bad and not adaptive to China.

\subsection{The adaptability of driverless technologies}
According to analyses above, we summarize the adaptability of AVs to pedestrians in Table \ref{table 13}. The interaction features could transfer the state of AVs to pedestrians but need a standard language to manage external interfaces. Driverless technologies could detect normal pedestrians well. Moreover, in small and distracted pedestrian detection, the precision is OK and there is room to promote. The worse scenes are pedestrians with heavy occlusion and detectors have very bad performance in the situation. Additionally, the influencing factors of receptivity reflects the importance of real experiences with AVs but little is conducted in China. Hence, the receptivity of Chinese pedestrians is so bad.

\begin{table*}[]
\centering
\caption{The adaptability of AVs to pedestrians in urban China}
\label{table 13}
\begin{tabular}{m{2.5cm}<{\centering}m{1.5cm}<{\centering}m{2cm}<{\centering}m{1.5cm}<{\centering}m{2cm}<{\centering}m{3cm}<{\centering}m{2.5cm}<{\centering}}\hline
\multirow{2}{*}{Technical demands}  & \multicolumn{4}{c}{AVs adapt to pedestrians: detection}          &\multicolumn{2}{c}{Pedestrians adpat to AVs}                                                         \\\cline{2-7}
                                                                   & Normal pedestrians & Occluded pedestrians         & Small pedestrians & Distrated pedestrians & {Interaction(HMI)} & {Receptivity}                            \\\hline
Evaluation                                                      & Excellent          & Bad                          & OK                & OK                   & OK  & Bad                          \\
Limitations and suggestions                                     & N/A                  &  Greatly improve precision in heavy occlusion(lowest miss rate:49\%) &  Improve detection precision(lowest miss rate:23\%)         &  More researches about phone distraction detection; more researches focusing on other distraction factors            & Call for standard HMI design language;avoid information overload of HMI; fuse different interface modalities
   & Totally lack receptivity research in China; conduct driverless experiments involving pedestrians                \\\hline
\end{tabular}
\end{table*}
      
\subsection{Influence on pedestrian mentality and government regulations.}
AVs would greatly change the traditional situation of pedestrian traffic and take a change to pedestrian mentality and regulations. In the scholar research, 
There are some articles talking about and the mentality of drivers in AVs and legal obstacles of pushing AVs into mass production, such as issue of compliance, issues of liability, issues of information governance\citep{RN181, RN284, RN285, RN182, RN283}. However, there is a lack of research on the change of pedestrian mentality and government regulations to pedestrians when AVs enter the road. Therefore, based on the analyzed behaviors of pedestrians and adaptability analysse, the AVs’ influence on pedestrian mentality and government regulations would be discussed in the following. \\
 Firstly, the mentality of pedestrians to conduct these bad behaviors would change. Currently, AVs are troubled by occluded pedestrian detection, which is to say, AVs would not pass the road when there are a few pedestrians. Furthermore, the human-machine interface would accurately transfer this state to pedestrians that AVs would stop. In the situation, pedestrians would get road rights with no fear of accidents and learn experience from it, which could cause a more serious red light running take place. Similarly, jaywalking would be safer for pedestrians because they could find the AVs are about to stop directly compared to gaining information from traditional drivers. As for distraction, pedestrians would have to pay attention to AVs because they are not familiar with this new technology and the human-machine interface should read by them themselves, which possibly reduces the distracted pedestrians. In summary, pedestrians would pay more attention to the traffic but turn bold to violate rules. \\
Under the context of AVs, pedestrian regulations should be adjusted accordingly. We conclude that pedestrians could be bolder to violate the rules above. In the situation, pedestrians would be safe but AVs would lose their ability to make traffic efficiency. In most of existing regulations, violated pedestrians are hard to caught and the punishment is light, which could not prevent bad behaviors in an efficient way. Therefore, the government must take stronger and more efficient measures to limit pedestrians violate the rules; for example, record illegal records into personal credit reports or build pedestrian bridges. Only by strong or efficient measures could reduce pedestrian violations and let AVs bring traffic more efficiency.

\section{Challenges \& Opportunities }
Based on the above analyses, we summarize some challenging but promising researches in the Chinese pedestrian environment.  
\subsection{Standard interaction language}
 As discussed above, a number of researches emerge on the external interfaces to interact AVs with pedestrians but there is not a common language. What information to transfer and how to transfer depends on researchers. We could imagine it will be a mess if different AVs use different interacting ways to inform pedestrians, causing a possible misunderstanding. Therefore, we call for a standard interacting language with two parts. Firstly, the kind of transferring information should be standardized. All AVs only transfer specific kinds of information, such as startup, observing the pedestrians, etc. Secondly, the information transferring methods should be made standard. For example, AVs would only indicate they are starting up by lighting red lights. Under the system, we believe pedestrians would have a better understanding of driverless technologies and push AVs to become real traffic participants in the long run.
\subsection{Interfaces fusion of HMI design}
Single modality interface, such as LED strips, might offer limited interaction information and have some shortages in its nature. Fusing interfaces of different modalities would be a promising solution, which would promote the precision and robustness of interaction. For example, pedestrians could observe all the interfaces and decide what to do. Especially, pedestrians could still judge the state of AVs when part of interfaces collapse, which is essential to protect pedestrian safety in the future with AVs. It is suggested that researchers should combine the advantages of different interfaces and eliminate the shortages by adding other interfaces. 
\subsection{Information overload avoidance on HMI design}
Interfaces fusion of HMI design has been a trend, however, overfull interfaces would lead to interaction information overload. In the situation, pedestrians tend to check all interafces and then decide what to do. This would confuse pedestrians and cause an inefficiency in traffic because pedestrians need much time to observe interfaces' state. Therefore, information overload on HMI design is suggested to avoid and the reriewed articles showed three interfaces are enough.
\subsection{Small pedestrian detection}
      Small pedestrian detection is important to protect safety around the world. In the Chinese driving environment, there are a number of small pedestrians due to jaywalking. However, the optimal detector only has a 23\% miss rate of small pedestrian detection, which is far from practical use. We suggest more efforts should be taken to improve the performance. 
\subsection{Heavily occluded pedestrian detection}
Occlusion often happens when pedestrians are crossing the crosswalk and is a challenging problem in the Chinese driving environment. Nevertheless, the detector performs poorly in heavily occluded pedestrians. Hence, we call on efforts to tackle the issues of heavily occluded pedestrian detection.             
\subsection{More distracted pedestrian research}
We have talked about the distracted pedestrian detection above and focus on smartphone usage. However, there is only one paper implemented to detect pedestrians using phones. Further research is required to push phone utilization detection. Moreover, other distracted reasons also call for researches in the context of AVs. Therefore, we suggest more distracted pedestrian research.      

\subsection{Nation-based receptivity research}
In the part of receptivity, we concluded factors that influence the receptivity of pedestrians, such as demography, knowledge and interaction, which highlight the real experiences between AVs and pedestrians. Therefore, the researches based on American would be not adaptive to other countries. However, as the largest market of cars, few researches are conducted on the receptivity in China. Therefore, we suggest nation-based receptivity research, especially in China, to get close knowledge to develop AVs. 
\subsection{Driverless experiments involving pedestrians. }
We conclude that real interacting experience of pedestrians greatly promote the trust to AVs. However, few researches are implemented involving pedestrians. To achieve receptivity, why not invite more pedestrians to join the empirical experiments? By doing so, not only do researchers gain practical feedback but also pedestrians could interact with AVs. In the process, pedestrians know more knowledge about AVs and hence promote the receptivity. Overall, it is wise for researchers to conduct experiments with pedestrians. 

\section{Conclusions}

The objective of this paper is to survey the adaptability of autonomous vehicles to pedestrians in urban China. China has a complicated pedestrian environment, however, little is known about the adaptability of future driverless technology. To this end, we analyzed three typical pedestrian behaviors in urban China and summed up the key technical demands for AVs. Then by reviewing the latest driverless technologies, we conclude the adaptability of autonomous vehicles to pedestrians respectively. Finally, we summarized the challenging problems and opportunities in Chinese pedestrian environment.\\
As we talked above, the adaptability of autonomous vehicles depends on the adaptability of interaction, detection as well as receptivity. In conclusion, driverless technologies perform well in normal pedestrian detection. Additionally, driverless technologies have an OK work in small pedestrian detection, distracted pedestrian detection and interaction. However, occluded pedestrian detection and the receptivity of pedestrians are not adaptive to China. 
We noted some challenging but promising areas for Chinese pedestrian environment, such as standard interaction languages and nation-based receptivity research. These aspects could be our future research or conducted by other practitioners.

\section*{Acknowledgment}
This research was funded by National Natural Science Foundation of China (Grant No. 51605054), State Key Laboratory of Vehicle NVH and Safety Technology (NVHSKL-202008 and NVHSKL-202010), The Science and Technology Research Program of Chongqing Education Commission of China (KJQN201800517 and KJQN201800107), Fundamental Research Funds for the Central Universities (No: 2019CDXYQC003), Chongqing Social Science Planning Project (No:2018QNJJ16), Key Technical Innovation Projects of Chongqing Artificial Intelligent Technology (cstc2017rgzn-zdyfX0039)

\bibliographystyle{cas-model2-names}
\bibliography{references}

\end{document}